\documentclass[aps,prx,twocolumn,superscriptaddress,10pt]{revtex4-2}
\usepackage[utf8]{inputenc}
\usepackage[english]{babel}
\usepackage[T1]{fontenc}
\usepackage{microtype}
\usepackage[pdftex]{graphicx}
\usepackage{hyperref}
\usepackage{amsmath}
\usepackage{amssymb}
\usepackage{braket}
\usepackage{xcolor}
\usepackage{dcolumn}
\usepackage{bbm}
\usepackage{bbold}
\usepackage[caption=false]{subfig}
\usepackage[capitalize]{cleveref}
\usepackage{float}
\usepackage{placeins}

\hypersetup{%
    hypertexnames=false,
    colorlinks=true,
    allcolors=blue,
    unicode=true
}

\newcommand{\phantomsubfloat}[1]{%
    {%
        \captionsetup[subfloat]{farskip=0pt,captionskip=0pt}
        \captionsetup[subfigure]{labelformat=empty}
        \subfloat{#1}
    }
}

\crefname{section}{Sec.}{Sec.}
\crefname{appendix}{App.}{App.}

\newcommand{\be}{\begin{equation}}
\newcommand{\ee}{\end{equation}}

\newcommand{\dd}{\text{d}}

\begin{document}

\title{Fast entangling gates for Rydberg atoms via resonant dipole-dipole interaction}
\author{Giuliano Giudici}
\affiliation{Institute for Theoretical Physics, University of Innsbruck, 6020 Innsbruck, Austria}
\affiliation{Institute for Quantum Optics and Quantum Information of the Austrian Academy of Sciences, 6020 Innsbruck, Austria}
\affiliation{PlanQC GmbH, 85748 Garching, Germany}
\author{Stefano Veroni}
\affiliation{PlanQC GmbH, 85748 Garching, Germany}
\author{Giacomo Giudice}
\affiliation{PlanQC GmbH, 85748 Garching, Germany}
\author{Hannes Pichler}
\affiliation{Institute for Theoretical Physics, University of Innsbruck, 6020 Innsbruck, Austria}
\affiliation{Institute for Quantum Optics and Quantum Information of the Austrian Academy of Sciences, 6020 Innsbruck, Austria}
\author{Johannes Zeiher}
\affiliation{Fakult\"at f\"ur Physik, Ludwig-Maximilians-Universit\"at M\"unchen, 80799 Munich, Germany}
\affiliation{Max-Planck-Institut f\"ur Quantenoptik, 85748 Garching, Germany}
\affiliation{Munich Center for Quantum Science and Technology (MCQST), 80799 Munich, Germany}

\begin{abstract}
The advent of digital neutral-atom quantum computers relies on the development of fast and robust protocols for high-fidelity quantum operations.
In this work, we introduce a novel scheme for entangling gates using four atomic levels per atom: a ground-state qubit and two Rydberg states.
A laser field couples the qubit to one of the two Rydberg states, while a microwave field drives transitions between the two Rydberg states, enabling a resonant dipole-dipole interaction between different atoms.
We show that controlled-Z gates can be realized in this scheme without requiring optical phase modulation and relying solely on a microwave field with time-dependent phase and amplitude. We demonstrate that such gates are faster and less sensitive to Rydberg decay than state-of-the-art Rydberg gates based on van der Waals interactions.
Moreover, we systematically stabilize our protocol against interatomic distance fluctuations and analyze its performance in realistic setups with rubidium or cesium atoms.
Our results open up new avenues to the use of microwave-driven dipolar interactions for quantum computation with neutral atoms.
\end{abstract}

\maketitle

\section{Introduction}
The last decade has witnessed the rapid development of quantum platforms based on Rydberg atom arrays. 
In addition to their remarkable success as analog quantum simulators~\cite{Bernien2017, Keesling2019, Ebadi2021, Semeghini2021}, neutral atoms trapped via optical tweezers or lattices are emerging as one of the most promising architectures for digital quantum computing, thanks to their scalability~\cite{tao2024, gyger2024, manetsch2024}, their long coherence times~\cite{Bluvstein2024, barnes2022, manetsch2024, graham2023, huie2023, norcia2023}, and their reconfigurable geometry that enables arbitrary qubit connectivity~\cite{bluvstein2022, Bluvstein2024}.
High-fidelity single-qubit and two-qubit quantum operations have also been demonstrated with several atomic species, including rubidium~\cite{levine2019,evered2023}, cesium~\cite{graham2022,infleqtion_gates2}, strontium~\cite{unnikrishnan2024,pucher2024,cao2024,finkelstein2024,tao2024}, and ytterbium~\cite{ma2022,ma2023,peper2024,atomcomputing_gates}. 
Yet, further improving their accuracy remains one of the outstanding challenges toward realizing a large-scale, fault-tolerant quantum computer with neutral atoms, making it crucial to develop novel schemes for Rydberg gates.

Typically, two-qubit gates with Rydberg atoms rely on the strongly repulsive van der Waals force arising when the two atoms are in the same Rydberg state~\cite{jaksch2000,levine2019,jandura2022,charles2023,evered2023,charles2024}. 
The state-of-the-art approach for these schemes involves state-selectively coupling the atomic qubit of each atom to one Rydberg level via a laser, whose amplitude and phase are modulated in time to yield a controlled-Z (CZ) gate up to a local phase~\cite{levine2019}.
The phase and amplitude pulses required to achieve the desired two-qubit gate are not unique, providing a degree of freedom that can be exploited to minimize gate execution time~\cite{jandura2022,pagano2022} or maximize gate robustness against fluctuations in specific parameters~\cite{charles2023,jandura2023}.
Time optimality is particularly relevant as the finite Rydberg lifetime is among the major sources of decoherence.
Such protocols have been tested in Rydberg arrays of rubidium~\cite{evered2023}, cesium~\cite{infleqtion_gates2}, strontium~\cite{tsai2024}, and ytterbium~\cite{peper2024,atomcomputing_gates} atoms, achieving gate fidelities above 99\%.

 \begin{figure}
     \includegraphics[width=\linewidth]{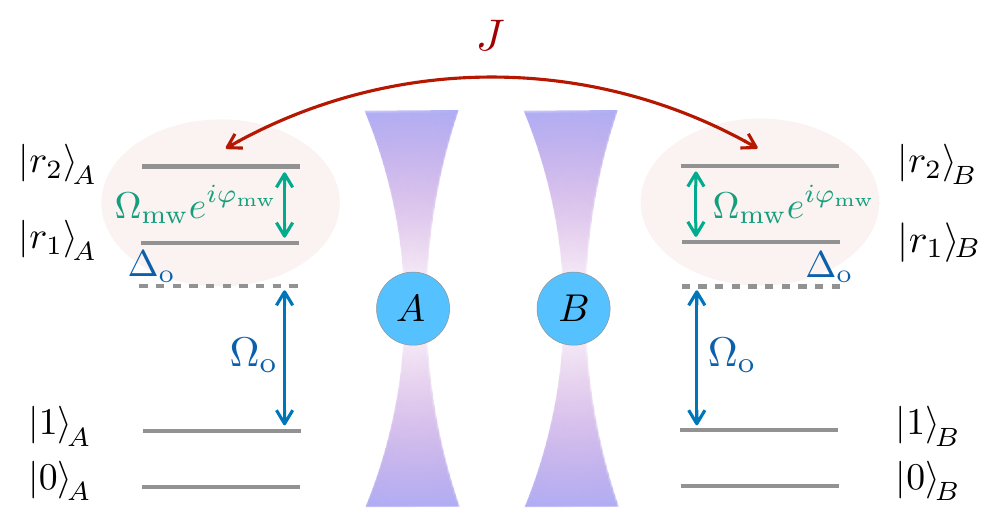}
     \vspace{-2mm}
    \caption{%
        Schematic representation of the level scheme utilized in this work for realizing a CZ gate up to a local phase.
        A laser field with amplitude $\Omega_\mathrm{o}$ and detuning $\Delta_\mathrm{o}$ couples the qubit state $\ket{1}$ to the Rydberg state $\ket{r_1}$.
        A microwave field with amplitude $\Omega_\mathrm{mw}$ and phase $\varphi_\mathrm{mw}$ couples the two Rydberg states $\ket{r_1}$ and $\ket{r_2}$ enabling a flip-flop interaction $J ( \ket{ r_1 r_2 } \! \bra{ r_2 r_1 } + \mathrm{H.c.} ) $ between the two atoms (cf. \cref{eq:ham_ideal}).
        Microwave field amplitude and phase are time-dependent control functions.
    }
    \label{fig1}
\end{figure}
\begin{figure}
     \phantomsubfloat{\label{fig2:a}}
     \phantomsubfloat{\label{fig2:b}}
     \includegraphics[width=\linewidth]{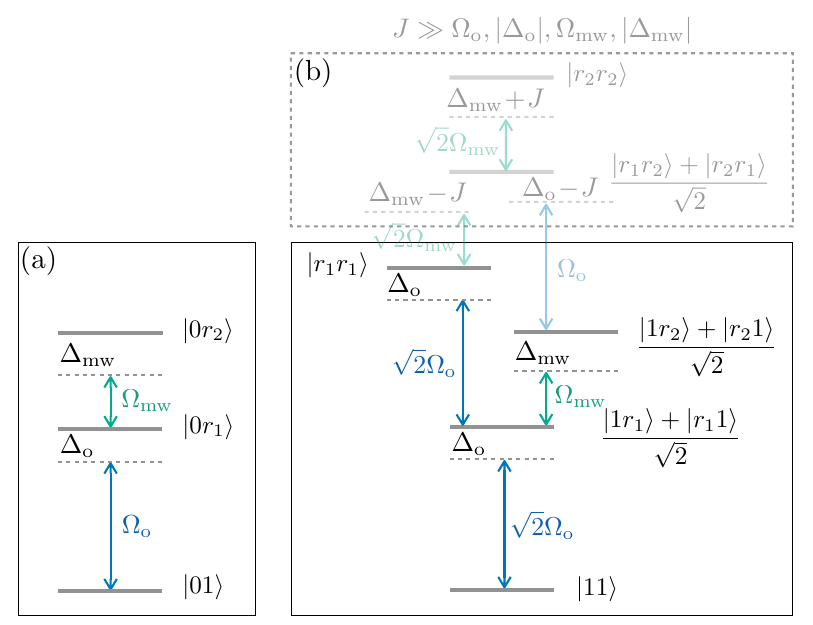}
     \vspace{-3mm}
    \caption{%
        Schematic representation of the two relevant blocks of the Hamiltonian \cref{eq:ham_ideal} that encode the dynamics of the state $\ket{01}$ (a) and $\ket{11}$ (b), after the unitary transformation $U = U_A \otimes U_B$, with $U_A = U_B = \mathrm{diag} ( 1 , 1, 1, e^{i \varphi_\mathrm{mw}})$ mapping the microwave phase to the detuning $\Delta_\mathrm{mw}  \! = \!  \frac{\dd \varphi_\mathrm{mw}}{\dd t}$.
        Because of the symmetry $A \leftrightarrow B$ of the protocol, all antisymmetric states are not relevant for the time evolution of $\ket{11}$. For $J \gg \Omega_\mathrm{o}, |\Delta_\mathrm{o}|, \Omega_\mathrm{mw}, |\Delta_\mathrm{mw}|$ the states in the shaded dashed box in panel (b) are decoupled from the dynamics of $\ket{11}$.
    }
    \label{fig2}
\end{figure}

Here, we present a different approach for realizing CZ Rydberg gates mediated by the dipole-dipole interaction between pairs of distinct Rydberg levels $\ket{r_1},\ket{r_2}$.
Our scheme is depicted in \cref{fig1}: the atomic qubit state $\ket{1}$ is optically coupled to the Rydberg state $\ket{r_1}$, and $\ket{r_1}$ is microwave coupled to nearby Rydberg level $\ket{r_2}$, enabling a flip-flop interaction of strength $J$ between the two atoms~\cite{henriet2020}.
Throughout the manuscript, we refer to $J$ as the resonant dipole-dipole interaction to distinguish it from the off-resonant, second-order van der Waals interaction commonly used in blockade-based gates.
A similar setup was considered in Ref.~\cite{yu2019} to construct adiabatic gate protocols.
The laser field can be configured as either global, to enable parallel gate operations, or local, to target individual pairs of atoms.
We use the microwave amplitude $\Omega_\mathrm{mw}$ and phase $\varphi_\mathrm{mw}$ as time-dependent control functions, and apply the Gradient Ascent Pulse Engineering (GRAPE) method~\cite{khaneja2005, garon2013, smith2013, anderson2015, riaz2019} to obtain the time-optimal protocol for this scheme.
We show that the resulting gate is up to $20 \%$ faster than the time-optimal ``van der Waals protocol'' put forward in Ref.~\cite{jandura2022}, while offering several other advantages such as lower sensitivity to Rydberg decay, increased interaction strengths for longer-range gates, and the possibility to combine optical addressability and global microwave control. Notably, our protocol relies solely on microwave phase and amplitude modulation and does not involve phase shaping of laser pulses, significantly simplifying hardware requirements and potentially reducing the impact of laser phase noise on gate fidelity.

We give a physical interpretation of the numerically obtained pulses in a regime where the resonant dipole-dipole interaction strength $J$ is much larger than all other energy scales, and find that the protocol can be formally interpreted as a reparametrization of the standard van der Waals protocol with time-dependent interaction strength.
We then focus on realistic experimental conditions and employ a GRAPE-based method for further optimizing the pulse shapes to make the gate robust against fluctuations of the interaction strength $J$ induced by the atomic motion.
Finally, we perform numerical simulations that take atomic motion and finite Rydberg lifetimes into account for rubidium and cesium atoms. The results indicate that Bell state fidelities exceeding $99.9 \%$ are achievable with our scheme, highlighting its feasibility and effectiveness.

The structure of the paper is as follows.
In \cref{sec2}, we discuss two different protocols for realizing entangling gates with resonant dipole-dipole interactions between Rydberg atoms in the limit $J/\Omega_\mathrm{o} \to \infty$.
In particular, the first one consists of two resonant laser $\pi$-pulses with a fast microwave pulse at constant detuning in between; the second requires a time modulation of the microwave detuning at constant laser detuning, and is mathematically equivalent to a van der Waals protocol with time-dependent interaction strength. 
In \cref{sec3}, we focus on the second protocol and lay out the numerical procedure employed to obtain the pulses that implement the CZ gate in the ideal case where no van der Waals interaction occurs between the Rydberg pair states.
In \cref{sec4}, we consider rubidium and cesium atoms where the Rydberg states $\ket{r_1}$ and $\ket{r_2}$ are, respectively, $P$ and $S$ states with principal quantum numbers from $n=40$ to $n=70$.
We compute their resonant dipole-dipole interaction strength $J$ and van der Waals interaction strengths $V_{11}, V_{12}, V_{22}$ at various distances, and repeat the GRAPE optimization to show that a qualitatively similar gate protocol exists in the presence of van der Waals forces. 
In \cref{sec5}, we optimize the pulse robustness against fluctuations of the interatomic distance and find a stabilized protocol that significantly enhances gate fidelities when considering the coupling to the atomic motional degrees of freedom.
We benchmark the performance of the robust pulses including both atomic motion and Rydberg decay, and demonstrate that our protocol competes with, and in some regimes even outperforms, state-of-the-art protocols.
In \cref{sec6}, we draw our conclusions and outline potential directions for future work.

\section{CZ gates from resonant dipole-dipole interaction}
\label{sec2}
The Hamiltonian that describes the two four-level systems depicted in \cref{fig1} reads 
\begin{align}
         \frac{H(t)}{\hbar} & =   \frac{\Omega_\mathrm{o} }{2} \left( \ket{1} \! \bra{r_1}_A +  \ket{1} \! \bra{r_1}_B + \mathrm{H.c.} \right) +  \nonumber \\[1mm]
&  -\Delta_\mathrm{o}  \left( \ket{r_1} \! \bra{r_1}_A +  \ket{r_1} \! \bra{r_1}_B  \right) +  \nonumber \\[1mm]
        & + \frac{\Omega_\mathrm{mw}(t)}{2}  \left[ e^{i \varphi_\mathrm{mw}(t)}  \left(  \ket{r_1} \! \bra{r_2}_A + \! \ket{r_1} \! \bra{r_2}_B \right) + \mathrm{H.c.} \right] + \nonumber \\[1mm]
        &  + J \left( \ket{ r_1 r_2 } \! \bra{ r_2 r_1 } + \mathrm{H.c.} \right),
        \label{eq:ham_ideal}
\end{align}
where $A$ and $B$ label the two atoms, $\Omega_\mathrm{o}$ and $\Delta_\mathrm{o}$ are the amplitude and detuning of the laser that couples the computational basis state $\ket{1}$ to the Rydberg state $\ket{r_1}$, $\Omega_\mathrm{mw}$ and $\varphi_\mathrm{mw}$ 
are the amplitude and phase of the microwave radiation that couples the two Rydberg states $\ket{r_1}$ and $\ket{r_2}$, and $J > 0$ is the strength of the dipolar exchange interaction.
Such interaction is present in Rydberg states for which a dipole transition is allowed, e.g. an $s$ state and a $p$ state (cf. \cref{sec4}).
The real Hamiltonian for the two atoms also includes van der Waals interaction terms of the form $V_{ij} \ket{r_i r_j} \! \bra{r_i r_j}$, which we will neglect in this section and discuss at length in \cref{sec4} and \cref{sec5}.

The most general maximally-entangling two-qubit gate realizable within this scheme is a CZ gate up to a single-qubit phase shift, which can be parameterized by an angle $\theta$ and can be written in the computational basis $\{ \ket{00},\ket{01},\ket{10},\ket{11} \}$ as
\begin{equation}
    \mathrm{CZ} (\theta) = \mathrm{diag} \left( 1 , e^{i \theta} , e^{i \theta} , -e^{i 2\theta} \right).
    \label{eq:CZ}
\end{equation}
We now outline two ways for obtaining such two-qubit gate from the dynamics \cref{eq:ham_ideal} when $J \gg \Omega_\mathrm{o}, |\Delta_\mathrm{o}|, \Omega_\mathrm{mw}, |\frac{ \dd \varphi_\mathrm{mw} }{ \dd t}|$.
We postpone the discussion of gate protocols at finite $J$ to \cref{sec3b}.
In what follows, we switch to a description in terms of the microwave detuning $\Delta_\mathrm{mw} = \frac{ \dd \varphi_\mathrm{mw} }{ \dd t} $ by applying the unitary transformation $U = U_A \otimes U_B$, where $U_A = U_B = \mathrm{diag} ( 1 , 1 , 1 ,e^{i  \varphi_\mathrm{mw} } )$.
The time evolution of $\ket{01}$ (or $\ket{10}$) and $\ket{11}$ is governed by the Hamiltonians $H_{01}$ and $H_{11}$ depicted in \cref{fig2}.
The flip-flop interaction only enters in $H_{11}$ and, for $J \gg \Omega_\mathrm{o}, |\Delta_\mathrm{o}|, \Omega_\mathrm{mw}, |\Delta_\mathrm{mw}|$, renders the states $\frac{1}{2} \left( \ket{r_1 r_2} + \ket{r_2 r_1} \right)$ and $\ket{r_2 r_2}$ far off-resonant and effectively decoupled from the dynamics. 
As a result, the optical Rabi frequency $\Omega_\mathrm{o}$ sets the gate timescale, and will fix our time units throughout this work.

A simple protocol for realizing a $\mathrm{CZ}(\theta)$ gate in this limit consists of three steps.
First a resonant optical $\pi$-pulse transfers the single qubit population from $\ket{1}$ to $\ket{r_1}$, such that
\begin{equation}
    U_1 \ket{01} = -i \ket{0 r_1}, \quad U_1 \ket{11} = -\ket{r_1 r_1}.
\end{equation}
A second off-resonant microwave pulse $U_2$ at constant detuning $\Delta_\mathrm{mw}$ is then applied to the atoms.
Since the laser is off during this pulse, the state $\ket{r_1 r_1}$ does not evolve under $U_2$ (cf. \cref{fig2}).
Hence, one can choose a pulse duration $T_\mathrm{mw} = 2 \pi / \sqrt{ \Delta_\mathrm{mw}^2 + \Omega_\mathrm{mw}^2 }$ such that $\ket{0 r_1}$ undergoes a complete Rabi oscillation $U_2 \ket{0 r_1} = e^{i \varphi_\mathrm{mw}} \ket{0 r_1}$, acquiring a phase $\pi \big( 1 + \Delta_\mathrm{mw} / \sqrt{\Delta_\mathrm{mw}^2 + \Omega^2_\mathrm{mw} } \big)$ that can be adjusted by tuning $\Delta_\mathrm{mw}$.
Setting $\Delta_\mathrm{mw} = \mp \Omega_\mathrm{mw} / \sqrt{3}$ yields $\varphi_\mathrm{mw} = \pm \pi/2$ and $\Omega_\mathrm{mw} T_\mathrm{mw} = \sqrt{3} \pi$.
The combined result for these two pulses is
\begin{equation}
    U_2 U_1 \ket{01} = -i e^{ \pm i \frac{\pi}{2} } \ket{ 0 r_1 }, \quad U_2 U_1 \ket{1 1} = -\ket{ r_1 r_1 }.
\end{equation}
Finally, a resonant optical $\pi$-pulse $U_3$ is applied to bring back the population to the computational basis:
\begin{equation}
    U_3 U_2 U_1 \ket{01} = -e^{ \pm i \frac{\pi}{2} } \ket{ 0 1 }, ~~~ U_3 U_2 U_1 \ket{1 1} = \ket{ 1 1 }.
\end{equation}
By comparing with \cref{eq:CZ}, one can see that this pulse sequence realizes a $\mathrm{CZ} ( \pm 3 \pi/2 )$ gate in a time
\begin{equation}
     \Omega_\mathrm{o} T =  2 \pi + \frac{ \sqrt{3} \pi \Omega_\mathrm{o} }{ \Omega_\mathrm{mw}} .
    \label{eq:gate_T1}
\end{equation}
As we discuss in \cref{app1}, when $\Omega_\mathrm{mw} \gg \Omega_\mathrm{o}$, the microwave pulse $U_2$ is much faster than the two laser pulses $U_1$ and $U_3$ and can be executed without turning off the laser with negligible effect on the gate fidelity.
This condition is easily achievable with standard microwave sources~\cite{pascal2022}.
Moreover, a proper time-modulation of the microwave phase can accommodate a finite resonant dipole-dipole interaction strength $J$.
However, we could not find a straightforward extension of this protocol that is stable upon the inclusion of van der Waals interactions $V_{ij}$.

A natural question that emerges at this point is whether both optical and microwave couplings can be used at the same time to find an improved gate protocol.
In the following, we present such a protocol and show that it can be systematically adapted to the case where van der Waals interactions are present and will be the focus of the rest of the main text.
We describe below its simplest version when $J \gg \Omega_\mathrm{o}, |\Delta_\mathrm{o}|, \Omega_\mathrm{mw}, |\Delta_\mathrm{mw}|$ and give its physical interpretation.
We assume laser and microwave fields to be always on with constant amplitudes $\Omega_\mathrm{o}$ and $\Omega_\mathrm{mw}$, and consider the microwave detuning $\Delta_\mathrm{mw} \gg  \Omega_\mathrm{o}, \Delta_\mathrm{o}, \Omega_\mathrm{mw}$.
In this limit the dynamics of $\ket{01}$ depicted in \cref{fig2} can be further restricted to only two levels. 
To see this, we diagonalize $H_{01}$ in the subspace $\{ \ket{0 r_1}, \ket{0 r_2} \}$ at first order in $\lambda = \Omega_\mathrm{mw}/2 \Delta_\mathrm{mw}$. $H_{01}$ expressed in the dressed basis $\{ \ket{01} , \ket{0 r_1} + \lambda \ket{0 r_2}, \ket{0 r_2} - \lambda \ket{0 r_1} \}$ reads
\begin{equation}
    \frac{H_{01}}{\hbar} = \! 
    \begin{pmatrix}
        0 & \frac{\Omega_\mathrm{o}}{2} & -\frac{ \lambda \Omega_\mathrm{o}}{2}  \\[1mm]
        \frac{\Omega_\mathrm{o}}{2} & -\Delta_\mathrm{o} \! + \! \lambda^2 \Delta_\mathrm{mw} &  0 \\[1mm]
        -\frac{ \lambda \Omega_\mathrm{o}}{2}  & 0 & -\Delta_\mathrm{o} \! -  \! \Delta_\mathrm{mw} ( 1 \! + \! \lambda^2 )
    \end{pmatrix}  .
\end{equation}
The AC Stark shift induced on the dressed states $\ket{0 r_1} + \lambda \ket{0 r_2}$ and $\ket{0 r_2} - \lambda \ket{0 r_1}$ can be made finite by setting $\Delta_\mathrm{mw} = \tau \Omega_\mathrm{mw}^2$.
If we now assume $\Omega_\mathrm{mw} \gg \Omega_\mathrm{o}, \Delta_\mathrm{o}$ the state $\ket{0 r_2}$ decouples from the dynamics of $\ket{01}$, which is governed by the two-level Hamiltonian 
\begin{equation}
    \frac{H_{01}}{\hbar} = \begin{pmatrix}
        0 & \frac{\Omega_\mathrm{o}}{2} \\[1mm] 
        \frac{\Omega_\mathrm{o}}{2} & -\Delta_\mathrm{o} + \frac{1}{4 \tau}
    \end{pmatrix}.
    \label{eq:H01}
\end{equation}
The same argument can be applied to $H_{11}$ upon replacing $\{ \ket{0 r_1},\ket{0 r_2}\}$ with $\{\frac{1}{\sqrt{2}} ( \ket{1 r_1} + \ket{r_1 1}  ),\frac{1}{\sqrt{2}} ( \ket{1 r_2} + \ket{r_2 1} ) \}$. In the basis $\{ 
 \ket{1 1} , \frac{1}{\sqrt{2}} ( \ket{1 r_1} + \ket{r_1 1}  ) , \ket{r_1 r_1} \}$ the resulting three-level Hamiltonian is
 \begin{equation}
 \frac{H_{11}}{\hbar} = 
\begin{pmatrix}
            0 & \frac{\Omega_\mathrm{o}}{\sqrt{2}} & 0  \\[1mm]
        \frac{\Omega_\mathrm{o}}{\sqrt{2}} & -\Delta_\mathrm{o} + \frac{1}{4 \tau} & \frac{\Omega_\mathrm{o}}{\sqrt{2}} \\[1mm] 
        0 & \frac{\Omega_\mathrm{o}}{\sqrt{2}} & - 2 \Delta_\mathrm{o}  
\end{pmatrix} .
    \label{eq:H11}
 \end{equation}
The dynamics described by \cref{eq:H01} and \cref{eq:H11} is mathematically equivalent to the one of two three-level systems $\{ \ket{0} , \ket{1} , \ket{r} \}$, where the state $\ket{rr}$ interacts  with van der Waals force $V = -1/ (2 \tau)$, and where the detuning from the Rydberg transition is given by $\Delta = \Delta_\mathrm{o} - 1/(4 \tau)$,
i.e. the two-qubit gate scheme of Ref.~\cite{levine2019}.
The Hamiltonian for this system is
\begin{align}
    \frac{H(t)}{\hbar} & =   \frac{\Omega_\mathrm{o} }{2} \left(  \ket{1} \! \bra{r}_A + \ket{1} \! \bra{r}_B + \mathrm{H.c.} \right) +  \nonumber \\[1mm]
    &   - \Delta(t) \left(  \ket{r} \! \bra{r}_A + \ket{r} \! \bra{r}_B  \right) + V(t)  \ket{ r r } \! \bra{ r r } .
 \label{eq:H_vdw_eff}
\end{align}
In the prototypical van der Waals gate, the distance between the atoms and thus the van der Waals coefficient $V$ are usually constant during the gate operation.
In our scheme, instead, a time-dependent $V$ simply corresponds to a time modulation of the microwave detuning $\Delta_\mathrm{mw}$. 
We will show in the next section that this additional control function provides a substantial speedup w.r.t. constant-$V$ van der Waals gates. 
In particular, we will demonstrate that the shortest gate time for a CZ operation with the Hamiltonian \cref{eq:H_vdw_eff} is $ T \simeq 6.03/\Omega_\mathrm{o}$, compared to the van der Waals gates execution time of $ T \simeq 7/\Omega_\mathrm{o}$ for $V/\Omega_\mathrm{o} \simeq 1.3$ (see \cref{app2}) and $T \simeq 7.6/\Omega_\mathrm{o} $ for $V/\Omega_\mathrm{o} = \infty$~\cite{jandura2022}.
We will also show that the obtained protocol can be extended to finite $\Omega_\mathrm{mw}/\Omega_\mathrm{o}$ and $J/\Omega_\mathrm{o}$ with only a slight increase of the gate execution time.

\section{Gate speed optimization}
\label{sec3}
In this section, we use GRAPE to find the time-optimal protocol that realizes a $\mathrm{CZ}(\theta)$ gate \cref{eq:CZ} within the scheme of \cref{fig1}.
This gate maps the product state $\ket{+ +}$ to the Bell state
\begin{equation}
    \ket{\psi_\theta } = \frac{1}{2} \left( \ket{00} + e^{i\theta} \ket{01} + e^{i\theta}\ket{10} - e^{2i\theta} \ket{11} \right) ,
    \label{eq:bell_state}
\end{equation}
such that the Bell state fidelity 
\begin{equation}
    F_\mathrm{Bell} = \left| \bra{ \psi_\theta } \mathcal{T} \exp \left( {-\frac{i}{\hbar} \int_0^T \!\!\!\! H(t) \, \dd t }  \right) \ket{ ++ } \right|^2 \!\! = 1,
\label{eq:bell_fid}
\end{equation}
where $\mathcal{T} \exp$ is the time-ordered exponential.
This quantity has been often used to benchmark and optimize the CZ gate~\cite{levine2019, saffman2021_recoil, theis2016, graham2019}. 
Since such gate is diagonal, $F_\mathrm{Bell} = 1 $ is equivalent to $F=1$, where $F$ is the average gate fidelity~\cite{pedersen2007}.
We take the Bell state infidelity $1 - F_\mathrm{Bell}$ as a cost function for the optimization, discretize time with a time step $\dd t$, and numerically minimize the cost function for a fixed total time $T$. 
The number of variational parameters for $\dd t = T/N$ is $k N + 2$, where $k$ is the number of control functions and $N$ is the number of time steps.
Such parameters are the values $f_i = f(t_i)$ of the unknown functions $f$ computed on the time-grid $t_i = ( i + \frac{1}{2} ) \dd t $, $i = 0,\dots N-1$, the detuning $\Delta_\mathrm{o}$ from the optical transition, and the single-qubit rotation angle $\theta$.
We will always keep a constant laser amplitude $\Omega_\mathrm{o}$ and constant detuning $\Delta_\mathrm{o}$. We use $\varphi_\mathrm{mw}(t)$ as a control function for $J/\Omega_\mathrm{o} = \infty$ and $\varphi_\mathrm{mw}(t), \Omega_\mathrm{mw}(t)$ for finite $J/\Omega_\mathrm{o}$.
We repeat the optimization for increasing $T$ until a time $T^*$ is found for which the cost function $\mathcal{C}$ vanishes.
The numerical minimization is performed using the method ``L-BFGS-B'' as implemented in the Python library SciPy~\cite{scipy}.
To speed up this procedure we provide the SciPy routine with the gradient of the (time-discrete) cost function, which can be straightforwardly computed analytically. Moreover, when the microwave amplitude $\Omega_\mathrm{mw}$ is also used as a variational control function, we set a lower bound $\Omega_\mathrm{mw}^\mathrm{min} = 0$ in the optimizer to ensure its positivity, and add a term to the cost function that enforces $\Omega_\mathrm{mw} (t_0) = \Omega_\mathrm{mw} (t_{N-1}) = 0$: 
\begin{equation}
    \mathcal{C} = 1 - F_\mathrm{Bell} + \left( \frac{ \Omega_\mathrm{mw} (t_0) }{ \Omega_\mathrm{o}}  \right)^2 + \left( \frac{ \Omega_\mathrm{mw} (t_{N-1}) }{ \Omega_\mathrm{o}}  \right)^2 .
    \label{eq:cost_f0}
\end{equation}

In the remainder of this section we employ GRAPE to first show that the scheme of \cref{fig1}, in the double limit $J/\Omega_\mathrm{o} \to \infty$ and $\Omega_\mathrm{mw}/\Omega_\mathrm{o} \to \infty$ with $\Delta_\mathrm{mw} = \tau \Omega_\mathrm{mw}^2$, enables the realization of an exact $\mathrm{CZ}(\theta)$ gate in a time $T \simeq 6.03/\Omega_\mathrm{o}$.
We then apply a modified version of the same method to find the time-optimal CZ gate protocol attainable with the Hamiltonian \cref{eq:ham_ideal} and a finite resonant dipole-dipole interaction strength $J$.

\subsection{Infinite \texorpdfstring{$J/\Omega_\mathrm{o}$}{JOm}}
\label{sec3a}
\begin{figure}
     \phantomsubfloat{\label{fig3:a}}
     \phantomsubfloat{\label{fig3:b}}
     \phantomsubfloat{\label{fig3:c}}
    \phantomsubfloat{\label{fig3:d}}
     \includegraphics[scale=0.45]{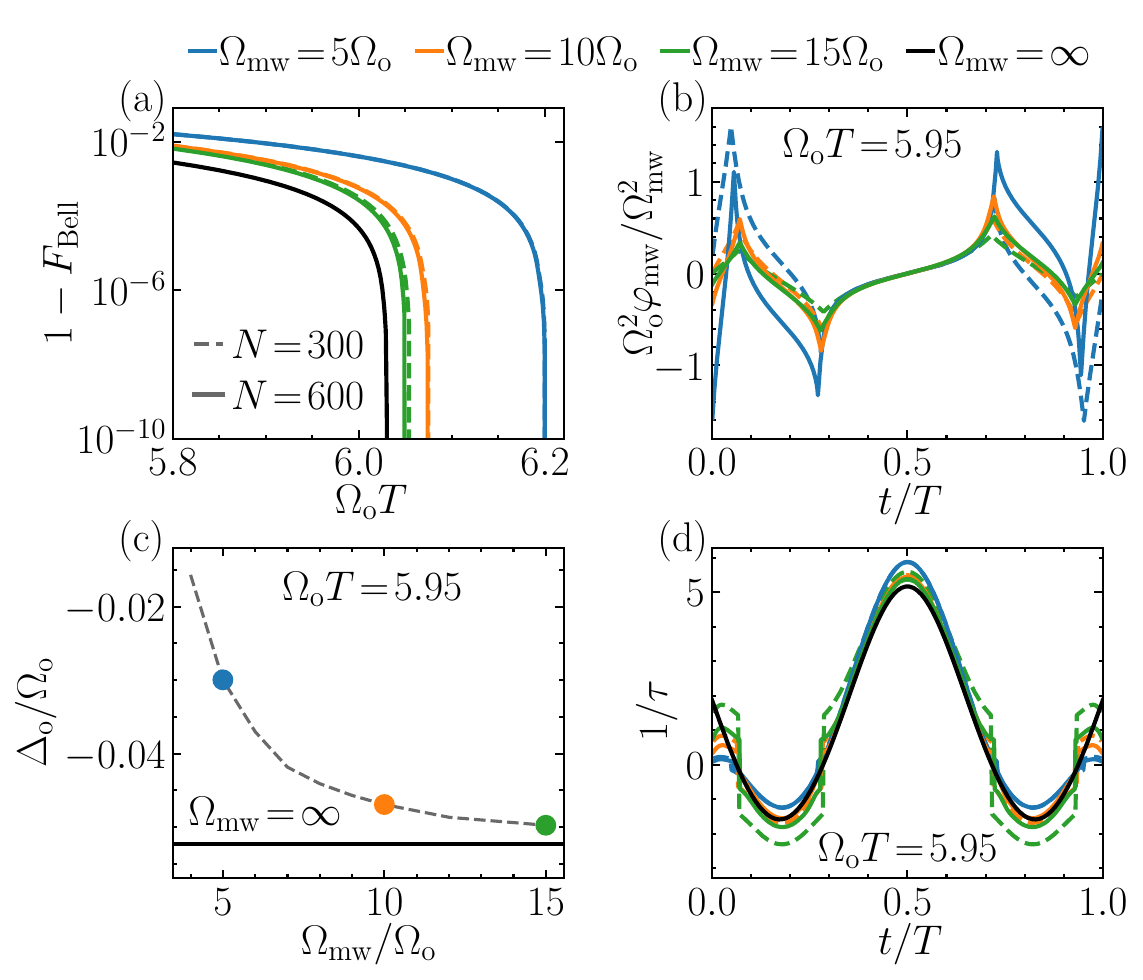}
     \vspace*{-5mm}
    \caption{%
        (a) Bell state infidelity as a function of the dimensionless gate time $\Omega_\mathrm{o} T$ obtained for the Hamiltonian \cref{eq:H_vdw_eff} (black lines) and for the Hamiltonian \cref{eq:ham_ideal} with $J/\Omega_\mathrm{o} = \infty$ for different values of $\Omega_\mathrm{mw}/\Omega_\mathrm{o}$ (colored lines).
        In the latter case, the optimization is performed for two different time step sizes $\dd t = T/N$, i.e.\ $N=300$ (dashed lines) and $N=600$ (solid lines), while in the former case $N=100$.
        (b) Optimal microwave phase obtained at finite $\Omega_\mathrm{mw}$ for different values of $\Omega_\mathrm{mw}/\Omega_\mathrm{o}$ and for $\Omega_\mathrm{o} T = 5.95$.
        (c) Optimal optical detuning $\Delta_\mathrm{o}$ for $\Omega_\mathrm{o} T = 5.95$ as a function of the microwave amplitude $\Omega_\mathrm{mw}$. For $\Omega_\mathrm{mw} / \Omega_\mathrm{o} \to \infty$, the result converges to the optimal $\Delta_\mathrm{o}$ obtained with the Hamiltonian \cref{eq:H_vdw_eff}.
        (d) Comparison between the pulse shapes optimized for the Hamiltonian \cref{eq:H_vdw_eff} (black lines) and for the Hamiltonian \cref{eq:ham_ideal} with $J/\Omega_\mathrm{o} = \infty$ (colored lines).
        The relation between the parameters of the two models is $\Omega_\mathrm{mw}^2 / \frac{\dd \varphi_\mathrm{mw}}{\dd t}  = -2 V \equiv 1/\tau$, and implies that $\varphi_\mathrm{mw}$ diverges when $V$ vanishes (cf.\ panel (b)).  
    }
    \label{fig3}
\end{figure}

In \cref{sec2}, we proved that the system of two dipole-dipole interacting four-level atoms depicted in \cref{fig1} maps to a system of two three-level atoms interacting via a time-dependent van der Waals force $V(t)$ and Rydberg-transition detuning $\Delta(t)$ \cref{eq:H_vdw_eff}.
This mapping is valid in the limits $J/\Omega_\mathrm{o} \to \infty$ and $\Omega_\mathrm{mw}/\Omega_\mathrm{o} \to \infty$ with $\Delta_\mathrm{mw} = \tau \Omega_\mathrm{mw}^2$, and the relations between the parameters of the two models are $V(t) = -1/(2 \tau(t))$ and $\Delta(t) = \Delta_\mathrm{o} -1/(4 \tau(t))$.
We now perform a GRAPE optimization on the Hamiltonian \cref{eq:H_vdw_eff} using $1/\tau(t)$, $\Delta_\mathrm{o}$ and the $\mathrm{CZ}$ gate angle $\theta$ (cf. \cref{eq:CZ}) as variational parameters.
The results of this optimization for $N=100$ time steps are the solid black lines in \cref{fig3}.
In \cref{fig3:a}, we plot the Bell state infidelity as a function of the dimensionless gate time $\Omega_\mathrm{o} T$.
The infidelity sharply drops to zero (within numerical precision) at the minimum time $T^*$ for which an exact CZ gate is realized, demonstrating the existence, under ideal conditions, of an exact $\mathrm{CZ}(\theta)$ gate in a time $T \simeq 6.03/\Omega_\mathrm{o}$.
In \cref{fig3:c}, we plot the time dependence of the control function $1/\tau(t) = -2 V(t)$ for $\Omega_\mathrm{o} T = 5.95$.
Such function has a direct interpretation in the resonant dipole-dipole interacting gate scheme of \cref{fig1} via $1/\tau(t) = \Omega_\mathrm{mw}^2/\frac{\dd \varphi_\mathrm{mw}}{\dd t}$. 

Although this correspondence only holds when $\Omega_\mathrm{mw}/\Omega_\mathrm{o} = \infty$, GRAPE remarkably finds a qualitatively similar solution for finite $\Omega_\mathrm{mw}/\Omega_\mathrm{o}$.
To show this, we minimize the Bell state infidelity obtained by evolving $\ket{++}$ with the Hamiltonian \cref{eq:ham_ideal}, using the microwave phase $\varphi_\mathrm{mw}(t)$ and optical detuning $\Delta_\mathrm{o}$ as variational parameters.
We enforce the constraint $J/\Omega_\mathrm{o} = \infty$ by projecting out the states $\ket{r_1 r_2}, \ket{r_2 r_1}, \ket{r_2 r_2}$ from the dynamics (cf. \cref{fig2:b}).
We plot in \cref{fig3:a} the minimum infidelity as a function of $\Omega_\mathrm{o} T$ for different values of $\Omega_\mathrm{mw}/\Omega_\mathrm{o}$, and in \cref{fig3:c} the function $1/\tau(t) = \Omega_\mathrm{mw}^2/\frac{\dd \varphi_\mathrm{mw}}{\dd t}$ for $\Omega_\mathrm{o} T = 5.95$, where $\frac{\dd}{\dd t}$ denotes the numerical derivative taken for a time step $T/N$ with $N=300,600$.
We compare this result to the one obtained from the Hamiltonian \cref{eq:H_vdw_eff} and observe a rapid convergence to the limit $\Omega_\mathrm{mw}/\Omega_\mathrm{o} = \infty$. 

The finite-$\Omega_\mathrm{mw}$ pulse depicted in \cref{fig3:d} is discontinuous for $t/T \simeq 0.06,0.28,0.72,0.94$.
This is because the optimal $V(t) \propto 1/\tau(t)$ for the model \cref{eq:H_vdw_eff} vanishes, causing divergences in the optimal $\varphi_\mathrm{mw}(t)$ for the model \cref{eq:ham_ideal} and slow convergence of the result with the number of time steps $N$ (cf.\ \cref{fig3:b,fig3:d}).
Although this discontinuity can be made arbitrarily small by reducing the time step $\dd t = T/N$, for too large $N$ the numerical optimization becomes unstable and eventually fails to produce a continuous solution.
This issue is even more severe when the constraint $J/\Omega_\mathrm{o} = \infty$ is relaxed. For this reason, for the GRAPE optimization at finite $J/\Omega_\mathrm{o}$ below we introduce a regularizer in the cost function that penalizes discontinuous solutions.

\begin{figure*}
    \centering
    \phantomsubfloat{\label{fig4:a}}
     \phantomsubfloat{\label{fig4:b}}
     \phantomsubfloat{\label{fig4:c}}
     \phantomsubfloat{\label{fig4:d}}
    \includegraphics[scale=0.45]{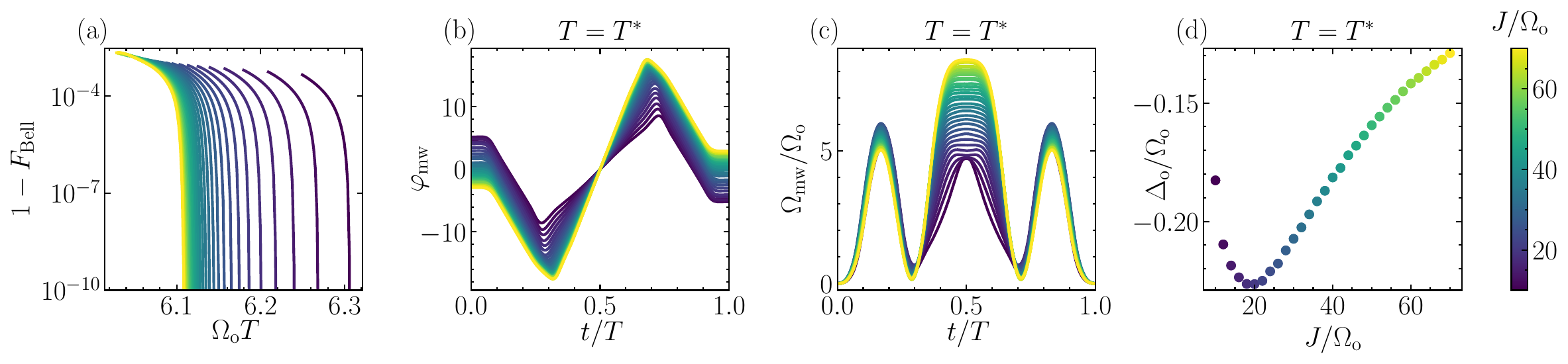}
    \vspace*{-5mm}
    \caption{%
        (a) Bell state infidelity as a function of the dimensionless gate time $\Omega_\mathrm{o} T$ for several values of $J/\Omega_\mathrm{o}$ ranging between $10$ and $70$ (cf.\ colorbar on the right of panel (d)).
        The number of time steps is set to $N=200$ and the regularizing parameter $\varepsilon = 10^{-3}$ (cf. \cref{eq:cost_f} and the text below).
        (b)--(d) Optimal microwave phase $\varphi_\mathrm{mw}$, microwave amplitude $\Omega_\mathrm{mw}$ and optical detuning $\Delta_\mathrm{o}$ at the time $T^*$ for which the time-optimal exact gate is found by the GRAPE optimization.
    } 
    \label{fig4}
\end{figure*}
\subsection{Finite \texorpdfstring{$J/\Omega_\mathrm{o}$}{JOm}}
\label{sec3b}
For the GRAPE optimization of the CZ gate protocol implemented via the Hamiltonian \cref{eq:ham_ideal} we use the control functions $\varphi_\mathrm{mw}(t), \Omega_\mathrm{mw}(t)$. 
As we show below, introducing the additional control function $\Omega_\mathrm{mw}(t)$ eliminates the divergences in the optimal microwave phase $\varphi_\mathrm{mw}(t)$ discussed in \cref{sec3a}.
Intuitively, this can be understood from the mapping of the infinite-$J$ limit of \cref{eq:ham_ideal} to the effective model of \cref{eq:H_vdw_eff}, where the diverging quantity is $\tau(t) = \frac{\dd \varphi_\mathrm{mw}}{\dd t} / \Omega_\mathrm{mw}^2$.
Therefore, divergences in $\varphi_\mathrm{mw}(t)$ can be avoided by allowing $\Omega_\mathrm{mw}(t)$ to vanish. 
To steer the optimization towards smooth solutions with vanishing $\Omega_\mathrm{mw}(t)$ rather than diverging $\varphi_\mathrm{mw}(t)$, we add a regularizer to the cost function that enforces the smoothness of the control fields

\begin{align}
       \mathcal{C}_\eta  & = \mathcal{C} + \eta \sum_{f}  \int_0^1  \left( \frac{ \dd f}{ \dd s} \right)^2  \dd s \nonumber \\ 
      & \simeq  \mathcal{C} + \eta\, N \sum_{f} \sum_{i=0}^{N-2} \left( f_{i+1} - f_i \right)^2, 
    \label{eq:cost_f}  
\end{align}
where $\mathcal{C}$ is the cost function \cref{eq:cost_f0}, $s = t/T$, $f = \varphi_\mathrm{mw}, \Omega_\mathrm{mw}$ and $\eta$ is a small constant that we adjust as the exact gate at time $T^*$ is approached.
Specifically, we initially set $\eta_0 = 10^{-6}$ for $T_0 < T^*$ and carry out the optimization until convergence.
We then take the resulting optimal pulses as initial conditions for the optimization at $T_1 = T_0 + \dd T$, with $\dd T =0.002 / \Omega_\mathrm{o} $, and reset $\eta_1 = \varepsilon \mathcal{C}_0$, where $\mathcal{C}_0$ is the Bell state infidelity obtained for the optimal solution at time $T_0$.
By iterating this procedure we systematically reduce $\eta$ as $T^*$ is approached.
We empirically find that the control functions obtained in this way are smooth and independent of the time step, as long as $\varepsilon \geq 10^{-3}$.
Increasing $\varepsilon$ results in a larger $T^*$.
Therefore, we tune $\varepsilon$ to the minimum value for which the optimal pulses are independent of the discretization scale $\dd t$. 

\begin{figure}
     \phantomsubfloat{\label{fig5:a}}
     \phantomsubfloat{\label{fig5:b}}
     \includegraphics[scale=0.45]{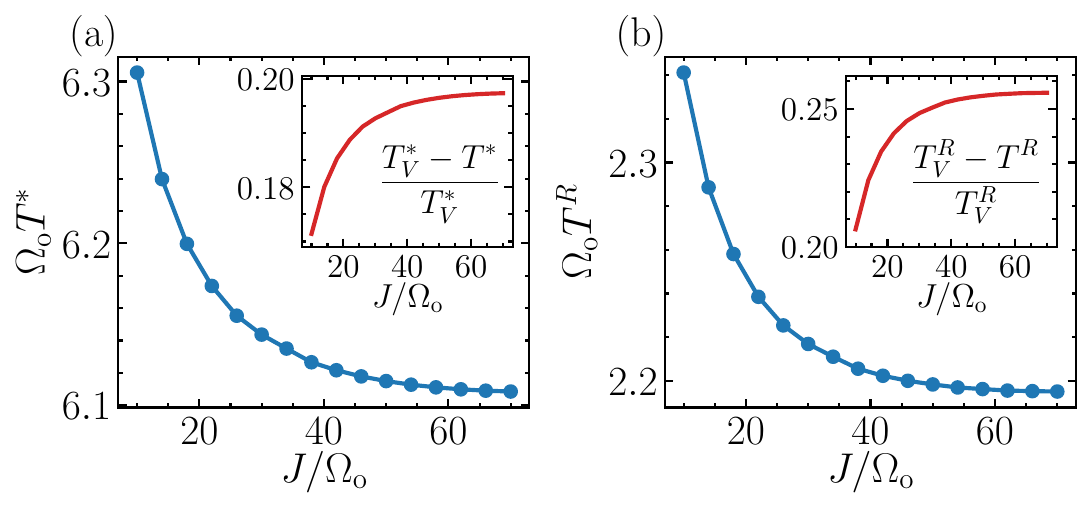}
     \vspace*{-2mm}
    \caption{%
        (a) Optimal CZ gate execution time for the pulses shown in \cref{fig4} as a function of $J/\Omega_\mathrm{o}$ (blue markers).
        The inset shows the relative speed-up $(T^*_V-T^*)/T^*_V$ w.r.t. the time-optimal van der Waals gate of Ref.~\cite{jandura2022} for which $\Omega_\mathrm{o} T^*_V \simeq 7.61$ (red line).
        (b) Same as (a) with the time spent in the Rydberg subspace $T^R$ \cref{eq:t_ryd} in place of the gate execution time $T^*$.
    }
     \label{fig5}
\end{figure}

The resulting pulses are plotted in \cref{fig4} for several values of $J/\Omega_\mathrm{o}$ and $N=200$.
In \cref{fig4:a} we show the Bell state infidelity as a function of the total time $T$.
The minimum gate execution time $T^*$ is $J$-dependent and decreases with increasing $J/\Omega_\mathrm{o}$.
As plotted in~\cref{fig5:a}, it ranges from $T^* \simeq 6.3/\Omega_\mathrm{o} $ for $J/\Omega_\mathrm{o} = 10$ to $T^* \simeq 6.1/\Omega_\mathrm{o} $ for $J/\Omega_\mathrm{o} = 70$. 
\Cref{fig5:a} also displays the relative speed-up with respect to the execution time $T^*_V \simeq 7.61/\Omega_\mathrm{o} $ of the time-optimal van der Waals gate~\cite{jandura2022}, ranging from 17\% up to 20\% for $J/\Omega_\mathrm{o} =10$ and $70$, respectively.
We observe that some of the divergences in the microwave phase $\varphi_\mathrm{mw}$ reported in \cref{fig3:b} become smooth peaks in \cref{fig4:c} at times $t/T \simeq 0.3, 0.7$ due to the regularizer.

Another important quantity to monitor is the time spent in the Rydberg manifold during the protocol, as it upper-bounds the infidelity due to the finite lifetime of the Rydberg states. 
It is given by
\begin{equation}
    T^R = \frac{1}{4}\sum_q \int_0^T \!\!\!\! \braket{q(t) | \left( \Pi_A + \Pi_B \right) | q(t) }\dd t,
    \label{eq:t_ryd}
\end{equation}
where $\Pi = \ket{r_1}\!\bra{r_1} + \ket{r_2}\!\bra{r_2}$ is the projector on the Rydberg subspace of one atom, the sum runs on all the computational basis states $\ket{q} = \left\{ \ket{00} , \ket{01} , \ket{10} , \ket{11} \right\}$, and $\ket{ q(t)}$ is the time evolution of these states under the optimal protocol.
We plot $\Omega_\mathrm{o} T^R$ vs $J/\Omega_\mathrm{o}$ in \cref{fig5:b}, demonstrating another substantial improvement ranging from 20\% to 26\% for $J/\Omega_\mathrm{o} = 10$ and $70$ with respect to the van der Waals gate $T^R_V \simeq 2.95/\Omega_\mathrm{o} $~\cite{jandura2022}.

\section{Implementation with alkali atoms}
\label{sec4}
So far, we neglected the van der Waals interactions arising when the two atoms are in a Rydberg state. Such interactions have the form
\begin{equation}
    H_\mathrm{vdW} = \sum_{i,j = 1 }^2 V_{i j} \ket{r_i r_j} \! \bra{r_i r_j } ,
    \label{eq:H_vdw}
\end{equation}
and have to be included in the Hamiltonian \cref{eq:ham_ideal}.
We do not expect the gate protocols depicted in \cref{fig3} to be sensitive to the value of the interaction strengths $V_{12}(=V_{21}),V_{22}$. 
In fact, $J \gg \Omega_\mathrm{o}$ and $|V_{12}|,|V_{22}| \gg \Omega_\mathrm{o}$ have the same effect on the state $\ket{11}$, decoupling the states $\ket{r_1 r_2},\ket{r_2 r_2}$ from its dynamics. On the contrary, $|V_{11}|$ has to be much smaller than $\Omega_\mathrm{o}$ to avoid the decoupling of $\ket{r_1 r_1}$, which plays an active role in our scheme, as we discussed in \cref{sec3}. Finally, we need to have $J \gg \Omega_\mathrm{o}$ since the execution time of our protocol decreases with increasing $J/\Omega_\mathrm{o}$ (cf. \cref{fig5:a}).
Combining these requirements we have the condition $J \gg \Omega_\mathrm{o} \gg |V_{11}|$.

The interactions strengths $J$ and $V_{11}$ depend on the interatomic distance $R$ as $J(R) \sim C_3/R^3$ and $V_{11}(R) \sim C_6/R^6$, respectively.
Hence, we can adjust $J/V_{11} \sim (C_3/C_6) R^3$ by tuning $R$.
However, $\Omega_\mathrm{o}$ has to be as large as possible since the real gate time $T \propto 1/\Omega_\mathrm{o}$. 
The optical Rabi frequencies achievable in typical experimental conditions $\Omega_\mathrm{o} /2 \pi \simeq   1-10 \, \mathrm{MHz}$ thus set the range of the required resonant dipole-dipole and van der Waals interactions to $J/2 \pi \simeq 10 - 100 \, \mathrm{MHz}$ and $ V/2 \pi \simeq 0.1 - 1 \, \mathrm{MHz}$. 
The $C_3$ and $C_6$ coefficients depend on the atomic species and the quantum numbers of the Rydberg states. Their scaling with the principal quantum number $n$ is $C_3 \sim n^4$ and $C_6 \sim n^{11}$~\cite{Adams2020}.
Therefore, $J/|V_{11}|$ decreases with $n$ for fixed $R$. 
Given the Rydberg lifetime scaling $\tau \sim n^2$ at room temperature, we need to find a tradeoff between large $J/|V_{11}|$ and long $\tau$. 
Below, we focus on rubidium with $n=40,50,60$ and cesium atoms with $n=40,50,60,70$, which are among the most used atomic species in Rydberg atom experiments~\cite{anand2024dualspecies, graham2023, Bluvstein2024,bornet2024}. 
In particular, we choose $\ket{r_1} = \ket{nP_{3/2},m_J = 3/2}$ and $\ket{r_2} = \ket{nS_{1/2},m_J = 1/2 }$. 
This choice is motivated by the fact that, in what follows, we will consider single-photon transitions from the computational qubit state $\ket{1}$---typically an $S$ state---to the Rydberg state $\ket{r_1}$, and a direct transition to an $S$ Rydberg state is forbidden by selection rules.
While for $n \geq 60$ the microwave transition between $\ket{r_1}$ and $\ket{r_2}$ falls in the $10-20\,\mathrm{GHz}$ range and can be directly modulated using routinely employed microwave techniques, for $n = 40$ and $n=50$ it lies in the more demanding $30-60\,\mathrm{GHz}$ range, which can be addressed using frequency modulation at a lower frequency~\cite{mw_modulation} combined with frequency multiplication or upconversion.
An alternative implementation based on a two-photon transition to $\ket{r_1}$ is discussed in \cref{app3}.

\begin{figure*}
     \phantomsubfloat{\label{fig6:a}}
     \phantomsubfloat{\label{fig6:b}}
     \phantomsubfloat{\label{fig6:c}}
     \phantomsubfloat{\label{fig6:d}}
    \phantomsubfloat{\label{fig6:e}}
     \phantomsubfloat{\label{fig6:f}}
     \phantomsubfloat{\label{fig6:g}}
     \phantomsubfloat{\label{fig6:h}}
     \center 
     \includegraphics[scale=0.45]{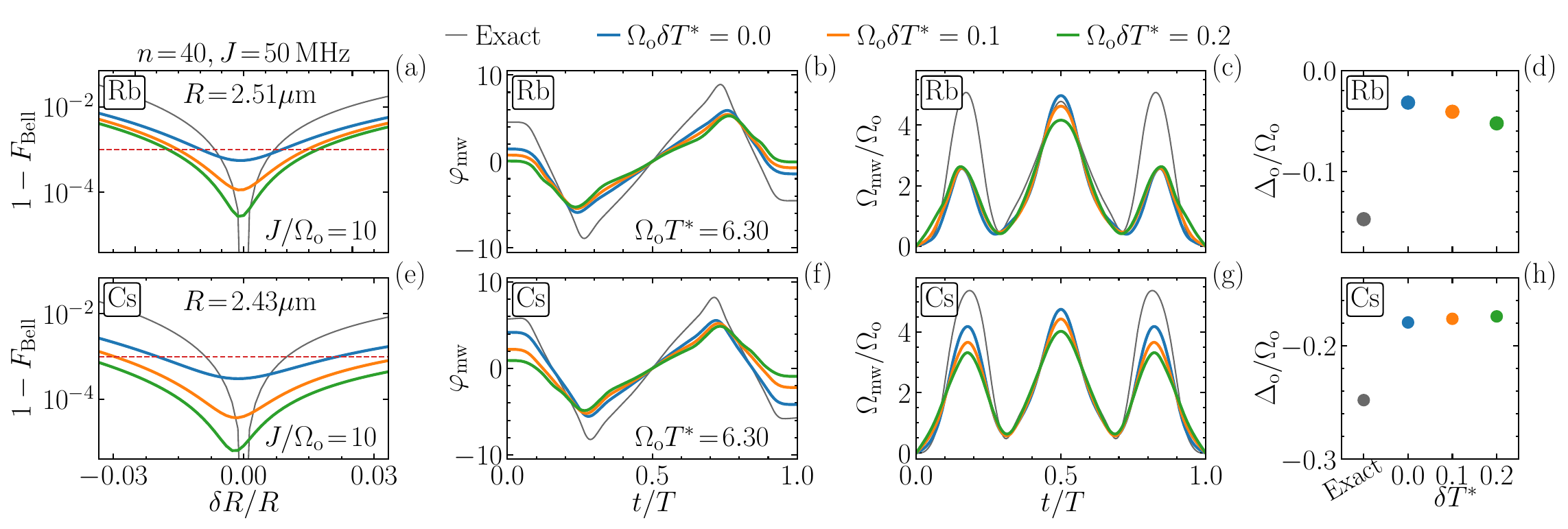}
     \vspace*{-2mm}
    \caption{%
        (a) Bell state infidelity as a function of the relative fluctuation in interatomic distance $\delta R/R$.
        The dark grey line corresponds to the time-optimal exact gate obtained with the procedure outlined in \cref{sec3b}.
        The colored lines correspond to the robust pulses obtained from the cost function \cref{eq:stability_f} with $\Omega_\mathrm{o} \delta T^* = 0,0.1,0.2$, where $\delta T^*$ is a slight increase of the time-optimal gate time $T^*$.
        Here $J/\Omega_\mathrm{o} = 10$ and the interaction parameters are in the first rows of \cref{tab} for rubidium (top) and cesium (bottom) Rydberg states.
        The horizontal red dashed line is $F_\mathrm{Bell} = 0.999$.
        The control functions are discretized on a time grid of $N=200$ points, while the integral in \cref{eq:cost_f} is discretized on $K=15$ points.
        The time-optimal pulses for this set of parameters have an execution time $T^* \simeq 6.30/\Omega_\mathrm{o}$.
        The pulse shapes for microwave phase $\varphi_\mathrm{mw}$ and amplitude $\Omega_\mathrm{mw}$ are plotted in panel (b) and (c), while the corresponding constant detuning $\Delta_\mathrm{o}$ are depicted in panel (d).
    }
     \label{fig6}
\end{figure*}

We calculate the interaction strengths $J,V_{ij}$ using the Python package \textsc{PairInteraction}~\cite{weber2017}. 
Selected results that will be employed in what follows are reported in \cref{tab}. We note that $V_{11} < 0$ for cesium, a feature that makes the gate more robust against fluctuations in $R$, as we will show in \cref{sec5} (cf. \cref{fig6:a}).

\begin{table}[t]
\setlength{\tabcolsep}{0.7em} 
\bgroup
\def\arraystretch{1.5}
\begin{tabular}{c c c c c c}
\multicolumn{6}{c}{ \bf{Rb}}\\
\hline
$n$ & $R \, (\mu \mathrm{m})$ & $J/2 \pi \, (\mathrm{MHz})$  & $V_{11}/J$ & $V_{12}/J$ & $V_{22}/J$  \\
\hline
40 & 2.51 & 50  & 0.007 & 0.016 & 0.079 \\
40 & 2.25 & 70  & 0.010 & 0.022 & 0.110 \\
50 & 3.45 & 50  & 0.015 & 0.037 & 0.182 \\
50 & 3.08 & 70  & 0.020 & 0.052 & 0.255 \\
60 & 4.45 & 50  & 0.027 & 0.076 & 0.354 \\
60 & 3.98 & 70  & 0.037 & 0.107 & 0.496 \\
\hline
\end{tabular}
\newline
\vspace*{3mm}
\newline
\begin{tabular}{c c c c c c}
\multicolumn{6}{c}{ \bf{Cs}}\\
\hline 
$n$ & $R \, (\mu \mathrm{m})$ & $J/2 \pi \, (\mathrm{MHz})$  & $V_{11}/J$ & $V_{12}/J$ & $V_{22}/J$  \\
\hline
40 & 2.43 & 50  & -0.011 & 0.007 & 0.064 \\
40 & 2.17 & 70  & -0.015 & 0.010 & 0.089 \\
50 & 3.36 & 50  & -0.021 & 0.020 & 0.150 \\
50 & 3.00 & 70  & -0.030 & 0.028 & 0.210 \\
60 & 4.35 & 50  & -0.038 & 0.045 & 0.295 \\
60 & 3.89 & 70  & -0.053 & 0.062 & 0.413 \\
70 & 5.41 & 50  & -0.062 & 0.087 & 0.514 \\
70 & 4.84 & 70  & -0.086 & 0.121 & 0.720 \\
\hline 
\end{tabular}
\egroup
\caption{%
    Resonant dipole-dipole and van der Waals interaction strengths $J$ and $V_{ij}$ between pairs of Rydberg states $\ket{r_1} = \ket{nP_{3/2},m_J = 3/2}$ and $\ket{r_2} = \ket{nS_{1/2},m_J = 1/2 }$ with $n=40,50,60$ for rubidium and $n=40,50,60,70$ for cesium atoms at a distance $R$.  
}
\label{tab}
\end{table}

We carried out the GRAPE optimization outlined in \cref{sec3b} including the van der Waals interaction \cref{eq:H_vdw} in the Hamiltonian \cref{eq:ham_ideal}, with microwave phase and amplitude as control functions. 
We set $8 \leq J/\Omega_\mathrm{o} \leq 14$ and $V_{ij}$ as listed in \cref{tab}. 
These interaction strengths enable Rabi frequencies $\Omega_\mathrm{o}/2 \pi \simeq 3.6 - 6.3 \, \mathrm{MHz}$ for $J/2 \pi = 50 \, \mathrm{MHz}$ and $\Omega_\mathrm{o}/2 \pi \simeq 5 - 8.75 \, \mathrm{MHz}$ for $J/2 \pi = 70 \, \mathrm{MHz}$. 
The time-optimal pulses resulting from this procedure are similar to the ones depicted in \cref{fig4} and yield almost the same gate time. One example obtained for $J/\Omega_\mathrm{o} = 10$ and the interaction strengths in the first row of \cref{tab} for rubidium and cesium Rydberg states are the pulses depicted in dark grey in \cref{fig6:b,fig6:c,fig6:d} and \cref{fig6:f,fig6:g,fig6:h}, respectively.
The gate total time and Rydberg subspace occupation time during the gate are also comparable to the values shown in \cref{fig5}, ranging from $\Omega_\mathrm{o} T \simeq 6.2$ and $\Omega_\mathrm{o} T^R \simeq 2.2$ for $J/\Omega_\mathrm{o} = 14$, to $\Omega_\mathrm{o} T \simeq 6.35$ and $\Omega_\mathrm{o} T^R \simeq 2.4$ for $J/\Omega_\mathrm{o} = 8$. 
\vspace{3em}

\section{Gate robustness optimization}
\label{sec5}

The main drawback of the experimentally realizable gate protocols of \cref{sec4} lies in the fact that the finite value of $J/\Omega_\mathrm{o}$ and the inclusion of small but non-negligible van der Waals forces make them sensitive to fluctuations $\delta R$ of the interatomic distance, which induce variations of the interaction strengths $\delta J/J = - 3 \, \delta R /R$ and $\delta V /V  = -6 \,\delta R / R$.
The same issue arises with standard van der Waals protocols away from the blockade regime.
To overcome this limitation we employ a simple cost function for GRAPE that targets pulse shapes more stable against small changes of $R$. 
Upon defining $x = \delta R/R$, such cost function has the form~\cite{khaneja2005}
\begin{align}
    \mathcal{C} & = 1 - \frac{1}{2 x_\mathrm{M}} \int^{x_\mathrm{M}}_{-x_\mathrm{M}}  \dd x \, F_\mathrm{Bell}( x ) ,
    \label{eq:stability_f} 
\end{align}
where $x_\mathrm{M} = \delta R_\mathrm{M}/R$ is the maximum fluctuation and $F_\mathrm{Bell} (x)$ is the Bell state fidelity corresponding to a Hamiltonian with $J \cdot  ( 1  -  3 x )$ and $V_{ij} \cdot  ( 1  - 6 x )$.

For the numerical optimization, we discretize the integral in \cref{eq:stability_f} over $K$ points and include the regularizer in \cref{eq:cost_f}, with $\eta = 10^{-7}$ to ensure the smoothness of the optimal control functions.
We use the time-optimal pulses obtained for a given set of parameters $J, V_{ij}$ as the initial condition for the robustness optimization and allow for a slight increase in the gate time, $T = T^* + \delta T^*$, where $T^*$ is the minimal time for realizing an exact $\mathrm{CZ}(\theta)$ gate (cf. \cref{fig4}).
This choice accelerates convergence, as the gradient evaluation in the robust cost function is more computationally intensive due to the averaging over atomic displacements $\delta R$.
We have verified that the final optimized pulses are robust with respect to perturbations in the initial guess and changes in the optimization hyperparameters $K$ and $x_\mathrm{M}$, consistently converging to the same control functions.
While we cannot exclude the existence of alternative robust protocols, possibly operating at shorter gate times, this appears unlikely given that $T^*$ typically sets a lower bound for the existence of exact or robust gates~\cite{jandura2023}.

The result of the GRAPE optimization is depicted in \cref{fig6} for the set of interaction strengths $J, V_{ij}$ in the first row of \cref{tab} for rubidium (top) and cesium (bottom) Rydberg states. While the stabilized pulses (colored lines) are qualitatively similar to the time-optimal exact protocol (dark grey lines), their Bell state infidelity in \cref{fig6:a} is one order of magnitude smaller over the whole interval $|\delta R / R| \leq 0.033$, excluding a small neighborhood around $\delta R / R = 0$. We also observe that the stabilized pulses for cesium yield lower infidelities over a broader range of $\delta R / R$, indicating that the attractive van der Waals interaction contributes positively to gate stabilization (cf. bottom and top of \cref{fig6:a}). The performance further improves when a small increase in the gate duration $\delta T^*$ is allowed.

\subsection{Gate performance with atomic motion and Rydberg decay}
\label{sec6a}

To benchmark the performance of the robust pulses such as those shown in \cref{fig6}, we carried out numerical simulations including the spontaneous Rydberg decay and the atomic motion induced by thermal fluctuations and photon recoil.
Below, we consider single-photon transitions to the Rydberg state $\ket{r_1}$. In such a setup, photon recoil and Rydberg decay account for most of the gate infidelity, provided laser phase and amplitude noise are negligible~\cite{tsai2024}. 
We note that single-photon excitation from hyperfine qubit states can also lead to unwanted coupling to multiple Rydberg Zeeman sublevels, even with pure light polarization.
Experimentally, this can be effectively mitigated by quickly transferring the qubit to a stretched hyperfine state prior to excitation, achievable with minimal infidelity~\cite{levine2022}, or by operating at large magnetic fields to spectroscopically isolate a single Rydberg transition.
Another possibility is to use a two-photon transition to a different Rydberg state $\ket{r_1}$, as discussed in detail in \cref{app3}.
We assume that the initial motional state is a thermal state $\rho \propto \exp \big( - \beta \hbar \omega_\mathrm{trap} \sum_{\ell = A,B} a_\ell^\dagger a^{\phantom{\dagger}}_\ell \big)$,
where $\omega_\mathrm{trap}$ is the trap frequency and $a$ ($a^\dagger$) are the annihilation (creation) operators of the relevant motional modes.
The tweezer traps are turned off during the gate, such that the vibrational modes' Hamiltonian reads
\begin{equation}
    H_{\mathrm{motion}} = \sum_{\ell = A,B } \frac{P_\ell^2}{2 m} \, , 
\end{equation}
where $P_\ell = i p_\mathrm{osc} ( a^\dagger_\ell - a^{\phantom{\dagger}}_\ell )$, $p_\mathrm{osc} = \sqrt{ \hbar  m \omega_\mathrm{trap} / 2  } $, and $m$ is the atomic mass.
The full model Hamiltonian is $H = H_{\mathrm{motion}} + H_{\mathrm{atom}} + H_{\mathrm{int}} + H_{\mathrm{decay}}$. 
The second term is the Hamiltonian of the atomic levels including the momentum transfer of the laser and microwave fields:
\begin{align}
             \frac{ H_\mathrm{atom}}{\hbar}  & =  \frac{\Omega_\mathrm{o}}{2}  \sum_{\ell = A,B } \! \left( e^{i \eta_\mathrm{o} ( a^{\phantom{\dagger}}_\ell + a^\dagger_\ell  ) } \ket{1} \! \bra{r_1}_\ell  + \mathrm{H.c.}  \right) + \nonumber \\[1mm] 
    & -\Delta_\mathrm{o}  \sum_{\ell = A,B } \!  \ket{r_1} \! \bra{r_1}_\ell + \nonumber \\[1mm] 
        & + \frac{\Omega_\mathrm{mw}}{2} \sum_{\ell = A,B } \left( e^{i \varphi_\mathrm{mw} + i \eta_\mathrm{mw}  ( a^{\phantom{\dagger}}_\ell + a^\dagger_\ell  ) } \ket{r_1} \! \bra{r_2}_\ell + \mathrm{H.c.} \right),
        \label{eq:recoil}
\end{align}
where $\eta_\mathrm{o} = 2 \pi x_\mathrm{osc} / \lambda_\mathrm{o}$ and $\eta_\mathrm{mw} = 2 \pi x_\mathrm{osc} / \lambda_\mathrm{mw}$ are the Lamb Dicke parameters for the optical and microwave transitions, with $x_\mathrm{osc} = \sqrt{ \hbar / ( 2 m  \omega_\mathrm{trap}  ) }$ and $\lambda_\mathrm{o} ( \lambda_\mathrm{mw}$) the optical (microwave) transition wavelength.
The third term is the interaction Hamiltonian where the resonant dipole-dipole and van der Waals potentials are expanded at first order in the interatomic distance fluctuations $\delta R = X_A - X_B = x_\mathrm{osc} \big( a^{\phantom{\dagger}}_A + a^\dagger_A - a^{\phantom{\dagger}}_B - a^\dagger_B \big)$:
\begin{align}
       \frac{ H_{\mathrm{int}} }{\hbar}  & = J \left( 1 -  \frac{3 ( X_A -  X_B )  }{R}  \right) \left(  \ket{r_1 r_2} \! \bra{r_2 r_1 } + \mathrm{H.c.} \right)  + \nonumber  \\[1mm]
       & + \sum_{i,j = 1}^2 V_{ij} \! \left( 1 -  \frac{6 ( X_A -  X_B ) }{R} \right)  \ket{r_i r_j} \! \bra{r_i r_j }. \label{eq:fluct_potentials}
\end{align}
Finally, the last term models the finite lifetimes of the Rydberg states $\ket{r_1},\ket{r_2}$ via a non-Hermitian Hamiltonian of the form
\begin{equation}
    \frac{H_\mathrm{decay}}{\hbar} = - i \sum_{j=1}^2 \frac{ \Gamma_j}{2} \left( \ket{r_j} \! \bra{r_j}_A + \ket{r_j} \! \bra{r_j}_B  \right),
    \label{eq:decay}
\end{equation}
where the lifetimes $1/\Gamma_i$ are taken from Ref.~\cite{lifetimes} and listed in \cref{tab2}.

\begin{table}
\setlength{\tabcolsep}{0.6em} 
\bgroup
\def\arraystretch{1.5}
\begin{tabular}{c c c}
\multicolumn{3}{c}{ \bf{Rb}}\\
\hline
$n$ & $\Gamma^{-1}_1  (\mu  \mathrm{s})$ & $\Gamma^{-1}_2 (\mu  \mathrm{s})$   \\
\hline
40 & 118  & 69 \\
50 & 239  & 141 \\
60 & 423  & 252 \\
\hline
\\
\end{tabular}\hspace{7mm}
\begin{tabular}{c c c}
\multicolumn{3}{c}{ \bf{Cs}}\\
\hline
$n$ & $\Gamma^{-1}_1 (\mu  \mathrm{s})$ & $\Gamma^{-1}_2  (\mu  \mathrm{s})$   \\
\hline
40 & 151  & 60 \\
50 & 313  & 126 \\
60 & 560  & 227 \\
70 & 913  & 372 \\
\hline
\end{tabular}
\egroup
\caption{%
Lifetimes for the Rydberg states $\ket{r_1} = \ket{nP_{3/2},m_J=3/2}$ and $\ket{r_2} = \ket{nS_{1/2},m_J = 1/2}$ employed for the numerical simulations presented in \cref{fig7} for rubidium (left) and cesium (right)~\cite{lifetimes}. \vspace{-1.5em}}
\label{tab2}
\end{table}

\begin{figure}
     \phantomsubfloat{\label{fig7:a}}
     \phantomsubfloat{\label{fig7:b}}
     \phantomsubfloat{\label{fig7:c}}
     \phantomsubfloat{\label{fig7:d}}
     \hspace*{-2mm}
     \includegraphics[scale=0.45]{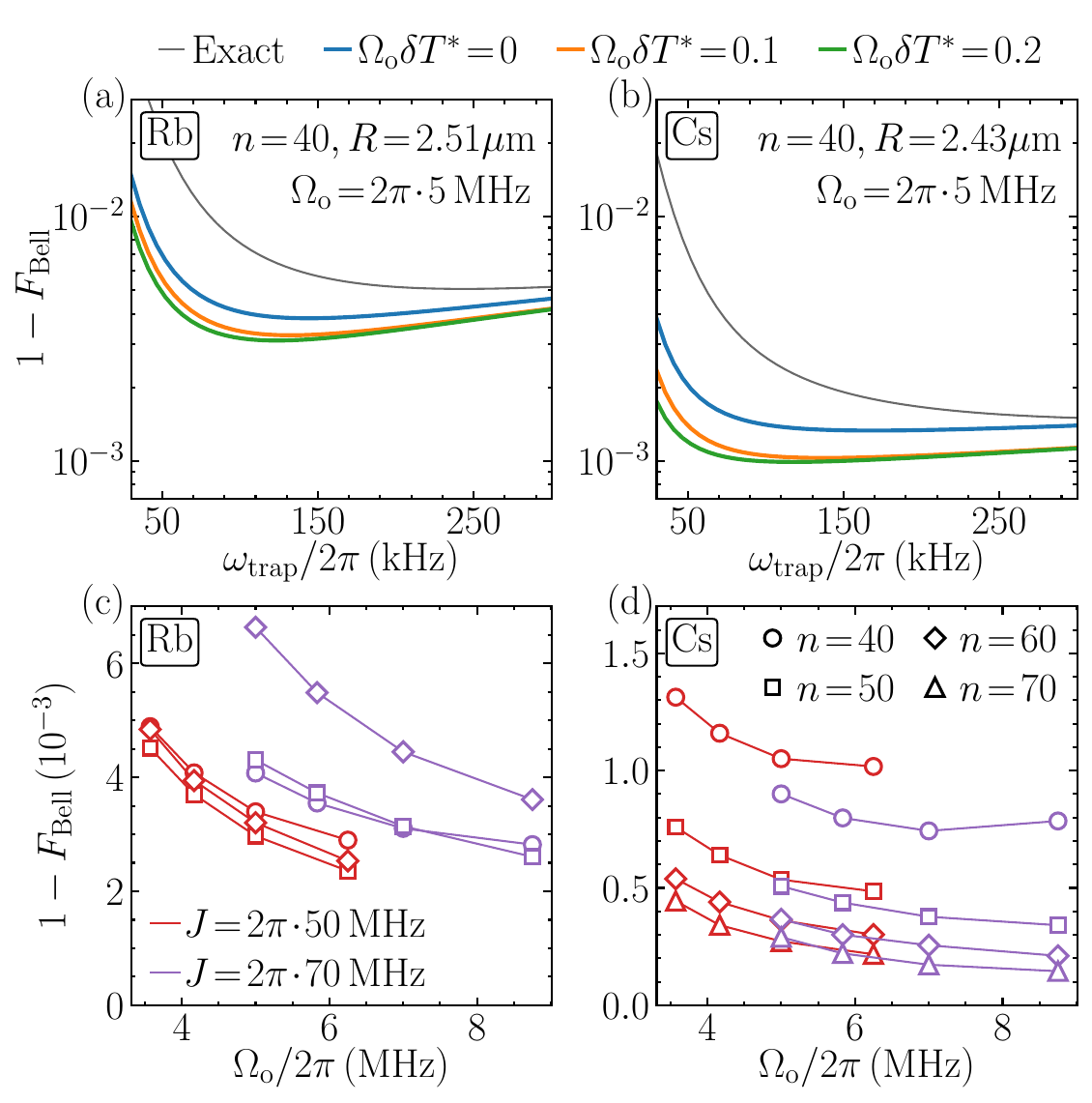}
     \vspace*{-2mm}
    \caption{%
        (a)--(b) Bell state infidelity due to atomic motion and Rydberg decay as a function of the trap frequency $\omega_\mathrm{trap}$ for the first row of the interaction parameters in \cref{tab} for rubidium (a) and cesium (b).
        The optical Rabi frequency is $\Omega_\mathrm{o}/2 \pi = 5 \, \mathrm{MHz}$.
        The dark grey line and the colored lines correspond to the exact time-optimal protocol and the robust protocols with a time increase $\delta T^*$, respectively.
        (c)--(d) Bell state infidelity as a function of the optical Rabi frequency $\Omega_\mathrm{o}$ obtained from the robust protocols with $\Omega_\mathrm{o} \delta T^* = 0.1$ for all the interaction parameters listed in \cref{tab} and a trap frequency $\omega_\mathrm{trap} / 2 \pi = 100 \, \mathrm{kHz}$.
        \vspace{-0.8em}
    }
    \label{fig7}
\end{figure}

For the numerical simulations, we set the initial temperature to $2 \, \mu \mathrm{K}$ and vary the trap frequency between 30 and 300 $\mathrm{kHz}$.
We used 8 vibrational modes per atom and verified that this number is sufficient to obtain converged results at all the considered trap frequencies.
The results are shown in \cref{fig7}.
We plot the Bell state infidelity as a function of $\omega_\mathrm{trap}$ in \cref{fig7:a} and \cref{fig7:b} for rubidium and cesium, respectively.
The values of $J$ and $V_{ij}$ are taken from the respective first rows of \cref{tab}.
The dark grey lines represent the exact protocol, whose execution time is $T^* \simeq 6.30/\Omega_\mathrm{o}$, while the colored lines are the robust protocols with $T = T^* + \delta T^*$ (cf. \cref{fig6}).
The latter considerably reduce the gate infidelity, especially at low trap frequencies.
Remarkably, the gate infidelity exhibits a non-monotonic dependence on the trap frequency, with an optimal $\omega_\mathrm{trap}$ that minimizes the error.
This behavior arises from two competing effects that dominate in different regimes. At low $\omega_\mathrm{trap}$, the dominant contribution to the infidelity arises from position fluctuations: in shallow traps, the large spatial extent of the atomic wavepacket makes the gate more sensitive to variations in interatomic distance.
As $\omega_\mathrm{trap}$ increases, the atoms become more tightly confined and this source of error is suppressed.
However, for sufficiently large trap frequencies, the infidelity increases again due to photon recoil: tighter traps reduce the spatial overlap between the kicked and unperturbed motional states, leading to increased decoherence, even when the atom remains in the motional ground state~\cite{saffman2021_recoil}.
We note that cesium yields the smallest infidelities thanks to its larger mass, which reduces the oscillator length $x_\mathrm{osc} \sim 1/\sqrt{m}$, its larger optical transition wavelength $\lambda_\mathrm{o}$, which reduces the photon recoil, and its attractive van der Waals force between $P$ states (see the discussion in \cref{sec6a}).

In \cref{fig7:c} and \cref{fig7:d}, we plot the gate infidelities obtained from robust protocols with $\Omega_\mathrm{o} \delta T^* = 0.1$ for all sets of interaction strengths in \cref{tab} and different optical Rabi frequencies $\Omega_\mathrm{o}/2 \pi \simeq 3.6 - 6.3 \, \mathrm{MHz}$ (red) and $\Omega_\mathrm{o}/2 \pi \simeq 5 - 8.75 \, \mathrm{MHz}$ (purple). 
The total gate times and Rydberg subspace occupation times for the stabilized protocols range from $T \simeq 6.45/\Omega_\mathrm{o}$ and $T^R \simeq 2.45/\Omega_\mathrm{o}$ to $T \simeq 6.30/\Omega_\mathrm{o}$ and $T^R \simeq 2.3/\Omega_\mathrm{o}$ for the largest and smallest Rabi frequencies, respectively.

We observe that cesium consistently yields lower gate infidelities—decreasing with increasing $n$—compared to rubidium across the range of parameters considered.
This behavior can be attributed to the nature of the van der Waals interaction. For cesium, $V_{11}/\Omega_\mathrm{o}$ is negative and increasingly attractive with $n$, which, similarly to large $J/\Omega_\mathrm{o}$, suppresses the population of the states $(\ket{r_1 r_2} + \ket{r_2 r_1})/\sqrt{2}$ and $\ket{r_2 r_2}$, thereby enhancing robustness to interatomic distance fluctuations.
At the same time, the Rydberg decay rate decreases with increasing $n$, leading to improved overall fidelity.
In contrast, for rubidium the van der Waals interaction is repulsive and increases with $n$, leading to a growing positive $V_{11}/\Omega_\mathrm{o}$.
This interaction suppresses the population of the state $\ket{r_1 r_1}$, which plays a key role in the dynamics underlying our gate protocol (see \cref{sec2}).
As a result, the positive $V_{11}/\Omega_\mathrm{o}$ competes with the beneficial effects of large $J/\Omega_\mathrm{o}$.
This competition reduces the robustness against interatomic distance fluctuations at large $n$, eventually making this the dominant source of infidelity despite the improved Rydberg lifetime.

\section{Conclusions and outlook}
\label{sec6}

We demonstrated how resonant dipole-dipole interactions between Rydberg atoms can mediate two-qubit entangling operations. Our proposed CZ gate protocols stand out by requiring only constant-amplitude laser pulses and time-modulated microwave fields, avoiding the need for complex control of optical phases and potentially suppressing the gate sensitivity to laser phase noise.
Compared to standard gate schemes based on van der Waals blockade, our approach is faster and less sensitive to the finite lifetimes of Rydberg states. Furthermore, we generalized our protocols to realistic atomic setups using rubidium and cesium atoms, and employed systematic stabilization methods to counteract fluctuations in atomic positions.
We showed that the stabilized protocols achieve Bell state fidelities on par with or exceeding current experimental realizations of neutral-atom entangling gates.

In this work, we only considered experimental implementations with heavy alkali atoms, for which reliable atomic physics calculations can be carried out.
Yet, alkaline-earth species such as strontium and ytterbium constitute promising candidates for realizing our scheme, respectively due to the attractive van der Waals interaction between Rydberg states in the singlet series for strontium, and the small predicted $C_6$ coefficients for states in the singlet series of ytterbium~\cite{vaillant2012}.
Another potential application of the gate protocol outlined in this paper is the realization of long-range gates, thanks to the slower decay with the distance of the resonant dipole-dipole potential with respect to to the commonly used van der Waals interaction. 
Finally, an interesting extension of this work is the generalization of our scheme to multi-qubit gates, which could natively be implemented on future neutral atom quantum computers.

\begin{acknowledgments}
We acknowledge useful discussions with Sebastian Blatt, Alexander Gl\"atzle, Andreas Kruckenhauser, Matteo Magoni, Cosimo C. Rusconi, Davide Dreon, Pascal Scholl, and Rick van Bijnen.
Giuliano Giudici acknowledges support from the European Union’s Horizon Europe program under the Marie Sk\l{}odowska Curie Action TOPORYD (Grant No. 101106005).
H.P. acknowledges support by the ERC Starting Starting Grant No.~101041435 (QARA), and the Austrian Science Fund (FWF): COE 1 and quantA.
J.Z. acknowledges support from the BMBF through the program “Quantum technologies - from basic research to market” (SNAQC, Grant No. 13N16265), from the Max Planck Society (MPG) the Deutsche Forschungsgemeinschaft (DFG, German Research Foundation) under Germany's Excellence Strategy--EXC-2111--390814868, from the Munich Quantum Valley initiative as part of the High-Tech Agenda Plus of the Bavarian State Government, and from the BMBF through the programs MUNIQC-Atoms. J.Z. is co-founder and shareholder of PlanQC GmbH.
This publication has also received funding under Horizon Europe programme HORIZON-CL4-2022-QUANTUM-02-SGA via the project 101113690 (PASQuanS2.1).
\end{acknowledgments}

\section*{Data Availability}
The data that support the findings of this article are openly available~\cite{ourdata}.

\appendix

\section{Piecewise protocol}
\label{app1}

As discussed in the main text, the piecewise protocol is composed of two optical $\pi$-pulses separated by a slightly-detuned microwave, with $\Delta_\mathrm{mw} = \mp \Omega_\mathrm{mw} / \sqrt{3}$, and it yields an exact $\mathrm{CZ}(\pm3 \pi / 2)$ gate in the limit of $J / \Omega_\mathrm{o} \to \infty$.
The time required is $\Omega_\mathrm{o} T =  2 \pi + \sqrt{3} \pi \,\Omega_\mathrm{o}/\Omega_\mathrm{mw}$, which, in the limit of large microwave driving ($\Omega_\mathrm{mw} \gg \Omega_\mathrm{o}$) yields a shorter time than the optimal van der Waals protocol.
The time spent in the Rydberg state, however, is slightly less favorable: $\Omega_\mathrm{o} T^R = \pi + \sqrt{3}\pi \Omega_\mathrm{o} / \Omega_\mathrm{mw}$.
Even in the large microwave driving limit, this is always larger than the van der Waals protocol ($\Omega_\mathrm{o} T^R \simeq 2.96$)~\cite{jandura2022}.

We also note that there is an additional solution for $\Delta_\mathrm{mw} \sim J$, which yields a slightly lower execution time of $\Omega_\mathrm{o} T =  2 \pi + \sqrt{2} \pi \,\Omega_\mathrm{o}/\Omega_\mathrm{mw}$.
Physically, this detuning brings the state $\ket{r_1 r_2} + \ket{r_2 r_1}$ into resonance with $\ket{r_1 r_1}$, as opposed to the previous case where all dynamics from $\ket{r_1 r_1}$ are trivial (cf. \cref{fig2}).
Similarly to the previous case, no phase modulation is necessary.

\begin{figure}
     \phantomsubfloat{\label{fig8:a}}
     \phantomsubfloat{\label{fig8:b}}
     \phantomsubfloat{\label{fig8:c}}
     \phantomsubfloat{\label{fig8:d}}
     \includegraphics[width=\linewidth]{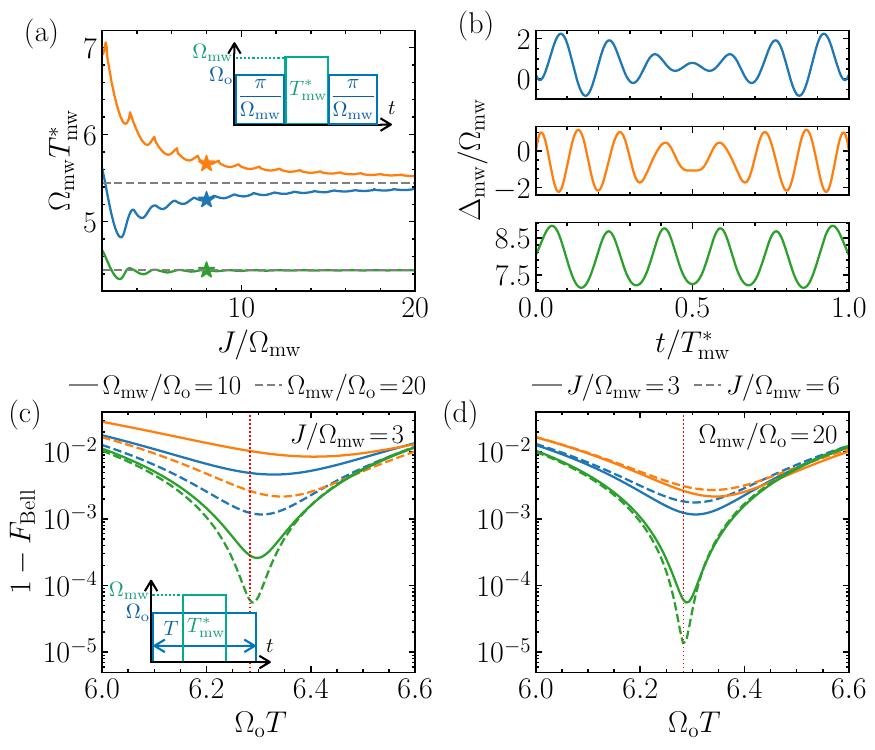}
     \vspace*{-5mm}
     \caption{%
        (a) Optimal pulse times for the different branches of solutions for the intermediate pulse in the exact piecewise protocol (see inset).
        The grey lines correspond to the asymptotic values $\sqrt{2} \pi$ and $\sqrt{3} \pi$.
        (b) Pulse shape for the different branches of solutions at $J / \Omega_\mathrm{mw} = 8$ (star).
        (c)--(d) Bell state fidelity for the approximate piecewise protocol without turning off the laser (see inset).
        The different branches from (a)--(b) are shown as a function of the duration of the laser pulse $\Omega_\mathrm{o} T$, for (c) fixed $J / \Omega_\mathrm{mw}$ and (d) fixed $\Omega_\mathrm{mw} / \Omega_\mathrm{o}$.
        As both parameters are increased, the optimal time $T^\ast$ approaches $2 \pi / \Omega_\mathrm{o}$ (red line).
    }
     \label{fig8}
\end{figure}

While the limit $J / \Omega_\mathrm{o} \to \infty$ is useful for gaining an analytical understanding, it is not the ideal regime for realizing fast gates in a practical setup.
For finite interaction strength, we resort to optimal control techniques.
Similarly to the main text, we use GRAPE to find an optimal modulation of the detuning $\Delta_\mathrm{mw}(t)$ for the intermediate pulse.
As shown in \cref{fig8:a}, feasible solutions for each branch are found up to $J / \Omega_\mathrm{mw} \simeq 2$.
The pulses consist of smooth oscillations around their asymptotic value (cf. \cref{fig8:b}), with a frequency increasing with $J / \Omega_\mathrm{mw}$.

It is worth noting that for large microwave driving, one can still realize an approximate $\mathrm{CZ}$ gate without turning off the optical drive during the intermediate pulse.
As shown in \cref{fig8:c,fig8:d}, by using the pulses found previously and simply adjusting the total laser pulse time $T$, one can achieve the desired gate with a fidelity improving both with $\Omega_\mathrm{mw} / \Omega_\mathrm{o}$ as well as $J / \Omega_\mathrm{o}$.
Notably, the branch with $\Delta_\mathrm{mw} \sim J$ seems to have the highest fidelities compared to the other two.
If the regime $J,\Omega_\mathrm{mw} \gg \Omega_\mathrm{o}$ is experimentally accessible, such a piecewise protocol can be appealing for practical implementations as it does not require any phase modulation on the laser beam.

\section{Finite-\texorpdfstring{$V$}{V} van der Waals protocols}
\label{app2}

In this section, we briefly discuss gate protocols using only a finite van der Waals interaction.
The setup is identical to Ref.~\cite{levine2019}.
We consider a single Rydberg state $\ket{r} = \ket{r_1}$ for each atom and the laser beam couples it to the computational state $\ket{1}$.
The total Hamiltonian is
\begin{align}
    \frac{H(t)}{\hbar} = & \frac{\Omega_\mathrm{o} (t) }{2} \left( e^{i \varphi_\mathrm{o}(t)} \ket{1} \! \bra{r}_A + e^{i \varphi_\mathrm{o}(t)} \ket{1} \! \bra{r}_B + \mathrm{H.c.} \right) + \nonumber \\[1mm]
    & + V \ket{r r}  \! \bra{r r},
    \label{eq:ham_blockade} 
\end{align}
Following the analysis of Ref.~\cite{levine2019}, this Hamiltonian splits into two blocks, each one becoming a two-level system in the limit of $V / \Omega_\mathrm{o}$ (the so-called blockade limit).
This was exploited to perform a pulse optimization using GRAPE, achieving the optimal time of $T \simeq 7.61 / \Omega_\mathrm{o}$~\cite{jandura2022}.

\begin{figure}
    \phantomsubfloat{\label{fig9:a}}
    \phantomsubfloat{\label{fig9:b}}
    \includegraphics[width=\linewidth]{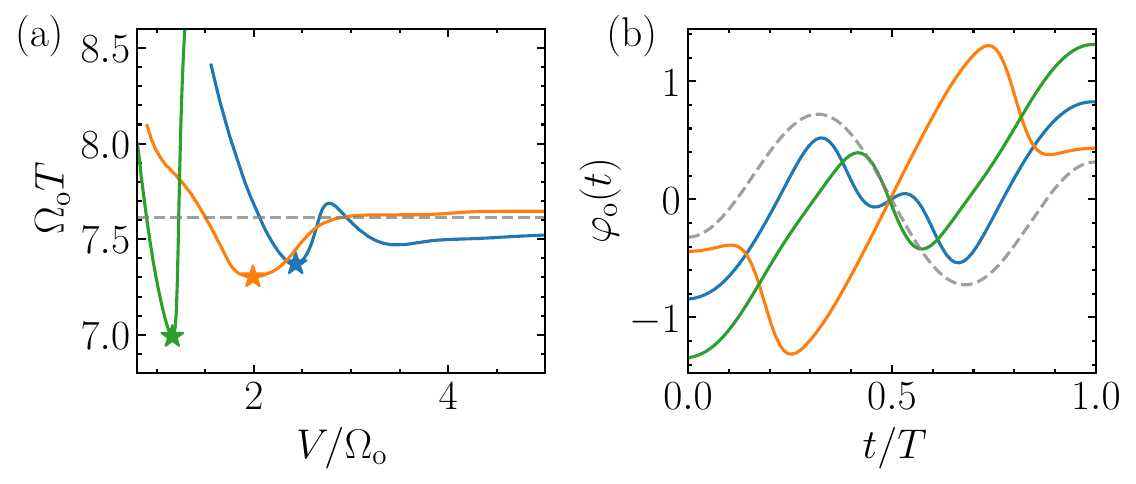}
    \vspace*{-5mm}
    \caption{
        (a) Optimal times for different sets of solutions at finite blockade strength $V$.
        The shortest pulse for each branch (highlighted with a star) is shown in (b).
        The dashed lines correspond to the values in the limit $V / \Omega_\mathrm{o} \to \infty$, previously found in Ref.~\cite{jandura2022}.
    }
    \label{fig9}
    \vspace*{-5mm}
\end{figure}

Performing a similar optimization for finite $V$, we obtain several sets of solutions based on the initial condition, as shown in \cref{fig9}.
Remarkably, numerically exact solutions can be found down to the regime of $V \simeq \Omega$.
We also note that, contrarily to the asymptotic case, the pulse time does not necessarily correlate with the total time in the Rydberg manifold $T^R$.
Similar pulses have been found in recent works designing Rydberg-dressed gate schemes~\cite{buchemmavari2023}.

\section{Gate protocol with two-photon Rydberg transition}
\label{app3}

\begin{table}
\setlength{\tabcolsep}{0.7em} 
\bgroup
\def\arraystretch{1.5}
\begin{tabular}{c c c c c c}
\multicolumn{6}{c}{ \bf{Rb}}\\
\hline 
$n$ & $R \, (\mu \mathrm{m})$ & $J/2 \pi \, (\mathrm{MHz})$  & $V_{11}/J$ & $V_{12}/J$ & $V_{22}/J$  \\
\hline
40 & 3.07 & 50  & 0.002 & 0.002 & -0.009 \\
50 & 4.20 & 50  & 0.004 & 0.005 & -0.108 \\
60 & 5.42 & 50  & 0.008 & 0.010 & -0.142 \\
70 & 6.70 & 50  & 0.014 & 0.018 & -0.207 \\
\hline 
\end{tabular}
\egroup
\caption{%
    Resonant dipole-dipole and van der Waals interaction strengths $J$ and $V_{ij}$ between pairs of rubidium Rydberg states $\ket{r_1} = \ket{(n-1)D_{5/2},m_J = 5/2}$ and $\ket{r_2} = \ket{nP_{3/2},m_J = 3/2 }$ with $n=40,50,60,70$ at a distance $R$.  
}
\label{tab3}
\end{table}

\begin{table}
\setlength{\tabcolsep}{0.6em} 
\bgroup
\def\arraystretch{1.5}
\begin{tabular}{c c c}
\multicolumn{3}{c}{ \bf{Rb}}\\
\hline
$n$ & $\Gamma^{-1}_1  (\mu  \mathrm{s})$ & $\Gamma^{-1}_2 (\mu  \mathrm{s})$   \\
\hline
40 & 55  & 118 \\
50 & 111  & 239 \\
\hline
\end{tabular}\hspace{7mm}
\begin{tabular}{c c c}
\multicolumn{3}{c}{ \bf{Rb}}\\
\hline
$n$ & $\Gamma^{-1}_1 (\mu  \mathrm{s})$ & $\Gamma^{-1}_2  (\mu  \mathrm{s})$   \\
\hline
60 & 196  & 423 \\
70 & 317  & 684 \\
\hline
\end{tabular}
\egroup
\caption{%
Lifetimes for the rubidium Rydberg states $\ket{r_1} = \ket{(n-1)D_{5/2},m_J=5/2}$ and $\ket{r_2} = \ket{nP_{3/2},m_J = 3/2}$ employed for the numerical simulations presented in \cref{fig10:d}~\cite{lifetimes}.}
\label{tab4}
\end{table}

In the main text, we analyzed implementations of our gate protocol based on a single-photon transition to the Rydberg manifold $\{ \ket{r_1} , \ket{r_2} \}$.
Here, we consider two-photon transitions in a setup with rubidium atoms analogous to that of Ref.~\cite{evered2023}.
Specifically, the Rydberg transition from the hyperfine qubit state $\ket{1}$ proceeds via the intermediate state $ \ket{e} = \ket{ 6 P_{3/2} } $. Since selection rules allow coupling to either $S$ or $D$ Rydberg states, we choose $\ket{r_1} = { (n-1) D_{5/2} , m_J=5/2}$ and $\ket{r_2} = { n P_{3/2} , m_J=3/2}$.
This choice is motivated by the significantly smaller ratio $V_{11}/J$ for $D$ states compared to $S$ states (cf. \cref{tab} and \cref{tab3}). 

\begin{figure}
     \phantomsubfloat{\label{fig10:a}}
     \phantomsubfloat{\label{fig10:b}}
     \phantomsubfloat{\label{fig10:c}}
     \phantomsubfloat{\label{fig10:d}}
     \includegraphics[scale=0.45]{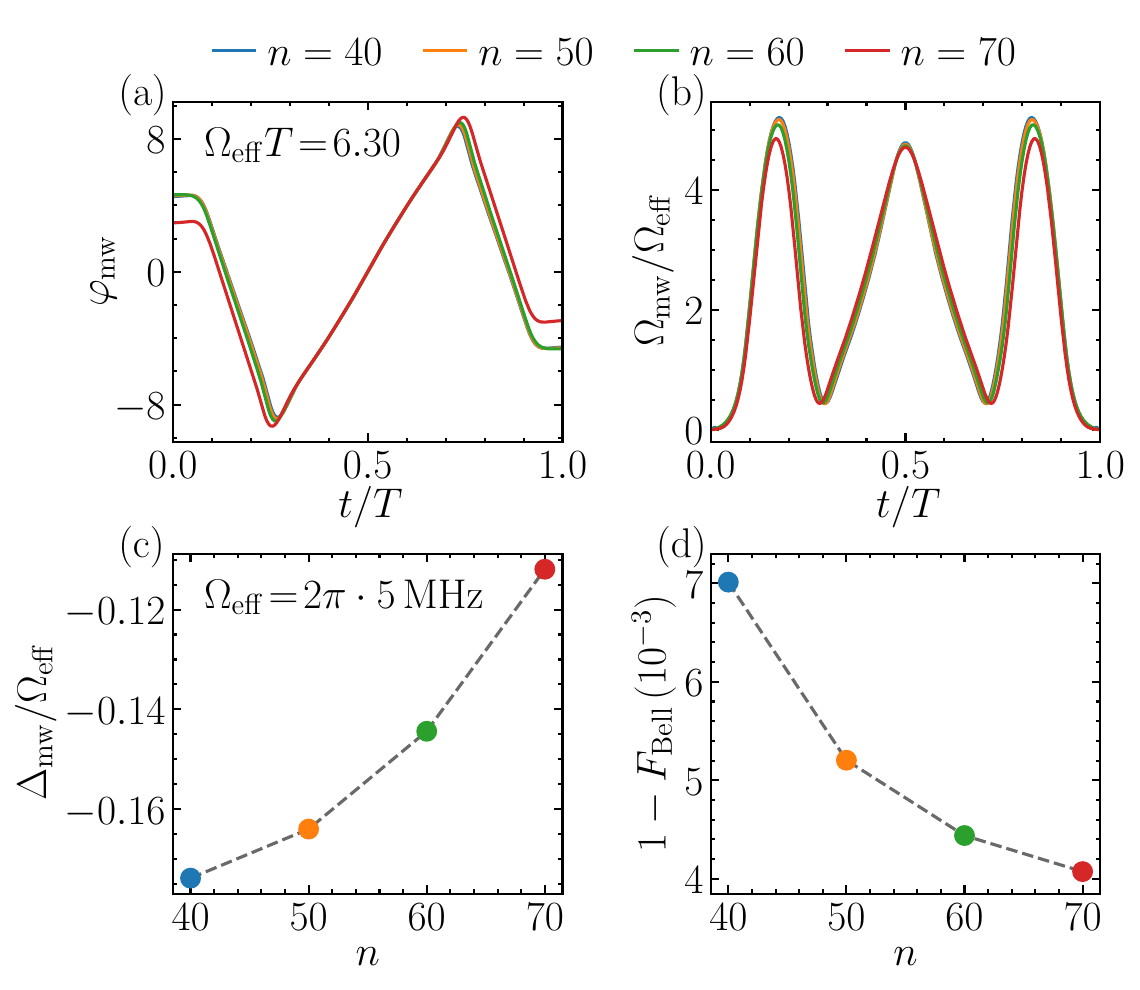}
     \vspace*{-5mm}
     \caption{%
     Optimal gate protocol for the rubidium Rydberg states in \cref{tab3}. The optimal microwave phase (a), microwave amplitude (b), and Rydberg detuning (c) are obtained upon adiabatic elimination of the intermediate state $\ket{e}$.
     (d) Gate infidelities computed under realistic experimental conditions with intermediate state detuning $\Delta_e  = 2\pi \cdot 7.75 \, \mathrm{GHz}$, and single-photon Rabi frequencies $\Omega_1 = \Omega_2 = 2\pi \cdot 278 \, \mathrm{MHz}$, which result in an effective two-photon Rabi frequency $\Omega_\mathrm{eff} = \Omega_1 \Omega_2 / 2 \Delta_e \simeq 2 \pi \cdot 5 \, \mathrm{MHz}$.
     The simulation also includes intermediate-state scattering, finite Rydberg lifetimes, and atomic motion with an initial motional state at $2 \, \mu \mathrm{K} $ and a trap frequency $\omega_\mathrm{trap} = 100 \, \mathrm{kHz}.$
    }
     \label{fig10}
\end{figure}

We optimize our gate protocol for rubidium Rydberg states with $n=40,50,60,70$, using the interaction parameters listed in \cref{tab2}. We carry out the GRAPE optimization assuming the adiabatic elimination of the intermediate state $\ket{e}$.
The resulting microwave phase $\varphi_\mathrm{mw}(t)$, amplitude $\Omega_\mathrm{mw}(t)$, and Rydberg detuning $\Delta_\mathrm{o}$ are shown in \cref{fig10:a,fig10:b,fig10:c}. Notably, the optimized phase and amplitude profiles are almost independent of $n$.
To refine the gate for finite intermediate state detuning $\Delta_e$, we reoptimize the single-qubit rotation angle $\theta$ (cf. \cref{eq:CZ}) and gate duration $T$ by maximizing the gate fidelity when $\ket{e}$ has a finite detuning $\Delta_e \gg \Omega_1, \Omega_2$, where $\Omega_1$ and $\Omega_2$ are the two single-photon Rabi frequencies.
We emphasize that the resulting gate protocol is not exact even in the idealized case as long as $\Delta_e/\Omega_1, \Delta_e/\Omega_2$ are finite. For $\Delta_e/\Omega_1 = \Delta_e/\Omega_2 = 27.8$, which we use in what follows, we obtain an ideal infidelity $1-F \simeq 0.001$.

We then benchmark the gate under realistic conditions, including the finite intermediate state lifetime $\tau_e = 110 \, \mathrm{ns}$, the finite Rydberg states lifetime listed in \cref{tab4}, and atomic motion. The Hamiltonian used for the numerical simulation is analogous to \cref{sec6a}, upon inclusion of the intermediate state $\ket{e}$ whose finite lifetime is modeled with an imaginary term analogous to \cref{eq:decay}. 
Similarly to \cref{sec6a} we take the motional degrees of freedom at an initial temperature of $2 \, \mu K$ with a trap frequency $\omega_\mathrm{trap}/2 \pi =  100 \, \mathrm{kHz}$ and assume the trap to be off during the gate. We set $\Omega_1/2\pi = \Omega_2/2\pi =  278 \, \mathrm{MHz}$ and $\Delta_e/2\pi = 7.75 \mathrm{GHz}$, such that effective two-photon Rabi frequency is $\Omega_\mathrm{eff} / 2 \pi \simeq 5 \mathrm{MHz}$. Following Ref.~\cite{evered2023}, we find that choosing $\Delta_e \Delta_\mathrm{o} < 0$ improves fidelities by reducing intermediate-state scattering. 
The resulting gate infidelity, plotted in \cref{fig10:d}, decreases with increasing $n$ from 0.7 \% for $n=40$ to $0.4 \%$ for $n=70$, thus demonstrating competitive performance relative to Ref.~\cite{evered2023}.

In contrast to \cref{sec5}, we did not find it advantageous to employ robust protocols such as those shown in \cref{fig6}.
We attribute this to the dominant role of intermediate-state decay and the intrinsic infidelity arising from finite $\Delta_e$.
A promising direction for further improvement would be to incorporate $\ket{e}$ explicitly into the GRAPE optimization, either to make the gate exact at finite $\Delta_e$ or to further suppress intermediate-state scattering.

\FloatBarrier
\bibliography{biblio.bib}
\end{document}